\documentclass[10pt,eqsecnum,twocolumn]{revtex4-1}
\usepackage[colorlinks=true]{hyperref}
\usepackage{verbatim}
\usepackage{graphicx} 
\usepackage{color} 
\usepackage{amsthm}
\usepackage{amsmath}
\usepackage{amssymb}
\usepackage{bbm,amsfonts}
\usepackage{setspace}
\usepackage{nomencl}
\usepackage{amstext} 

\newcommand{\Eop}{\mathcal{E}}

\newcommand{\Id}{\mathbbm{1}}

\newcommand{\braa}{\langle \psi |}
\newcommand{\kett}{|\psi\rangle}

\newcommand{\Hilb}{\mathcal{H}}

\newcommand{\tr}{{\mathrm{tr}}}

\newcommand{\gatefil}{\mathcal{F}_{\mathcal{E},\mathcal{U}}}

\newcommand{\FEM}{\overline{F_{\Eop,\mathcal{M}}}}
\newcommand{\rEM}{r_{\Eop,\mathcal{M}}}

\newcommand{\Meas}{\mathcal{M}}

\def\>{\rangle}
\def\<{\langle}

\begin{document}

\title{Experimentally efficient methods for estimating the performance of quantum measurements}

\author{Easwar Magesan}
\affiliation{Research Laboratory of Electronics and Nuclear Science and Engineering Department, MIT, Cambridge MA, 02139, U.S.A.}
\author{Paola Cappellaro}
\affiliation{Research Laboratory of Electronics and Nuclear Science and Engineering Department, MIT, Cambridge MA, 02139, U.S.A.}

\begin{abstract}
Efficient methods for characterizing the performance of quantum measurements are important in the experimental quantum sciences. Ideally, one requires both a physically relevant distinguishability measure between measurement operations and a well-defined experimental procedure for estimating the distinguishability measure. Here, we propose the average measurement fidelity and error between quantum measurements as distinguishability measures. We present protocols for obtaining bounds on these quantities that are both estimable using experimentally accessible quantities and scalable in the size of the quantum system. 
We explain why the bounds should be valid in large generality and illustrate the method via numerical examples.
\end{abstract}

\maketitle


\section{Introduction}\label{sec:introduction}

Measurement plays a fundamental role in the quantum sciences as it allows one to evolve and extract information from a quantum system. From a practical perspective, measurements are imperative for the successful implementation of a wide variety of protocols in quantum information processing (QIP) and communication~\cite{Fey82}, as well as high-precision metrology~\cite{GLM}. If large-scale QIP is achieved experimentally, certain tasks that are conjectured to be classically inefficient will become realizable using a scalable number of resources~\cite{Sho94,Llo96}. 

Various models of computation have been proposed for QIP, such as the standard circuit model~\cite{Deutsch85}, measurement based quantum computation (MBQC)~\cite{RB01}, and topological quantum computation~\cite{Kit97-2}. Each of these models relies greatly on the ability to perform accurate measurements. A large number of physical systems have been proposed as candidates for implementing QIP. A short, and certainly non-exhaustive, list of systems includes superconducting circuits~\cite{BVJED,KYG}, Nitrogen-vacancy (NV) centers~\cite{WJ06}, trapped ions~\cite{CZ}, NMR~\cite{CFH}, quantum dots~\cite{LD}, and optical implementations~\cite{KLM}. Measurement schemes vary greatly across these systems and so it is important to have straightforward protocols for comparing measurements that are both independent of the particular type of implementation and scalable in the size of the system. Providing such protocols is the main goal of this paper.

There has  been a significant amount of research towards \emph{completely characterizing} the error on the operations used to process and measure quantum information via quantum process~\cite{CN,ABJ,ML06} and measurement tomography~\cite{LFC,ZCD}. In principle, these tomographic methods can be used to characterize the error affecting any quantum operation or measurement realized in an experimental setting. Unfortunately, there are various significant drawbacks to complete process and measurement tomography. Process tomography requires an exponential number of resources in the number $n$ of quantum bits (qubits) that comprise the system (the number of parameters required to just describe the process scales as $16^n$), is not robust against state-preparation and measurement errors~\cite{MGS}, and requires intensive classical post-processing of measurement data~\cite{MSB}. Not surprisingly, since measurement tomography is realized by essentially inverting process tomography~\cite{LFC,ZCD}, it suffers from similar drawbacks to quantum process tomography. In particular, complete measurement tomography requires the ability to prepare a complicated set of pure input states with extremely high precision (possibly by performing complex unitary gates), scales badly in the size of the system, and can require lengthy post-processing of the tomographic data.

In many cases, one may only be interested in a subset of parameters characterizing the noise, or determining the \emph{strength} of the noise process rather than the entire process itself. For instance, in fault-tolerance~\cite{Sho96,AB-O,KLZ,Pre97}, as long as the strength of the noise affecting state-preparation, gates, and measurements is bounded by some value, large-scale computation is possible. As a result, various methods for \emph{partially characterizing} the error of quantum gates~\cite{ESMR,SMKE,KLRB,MGE,MGJ,MSRL,BPP,SLCP,SBLP,FL} have recently been proposed which attempt to circumvent many of the problems associated with full process tomography. The error is defined via the ``quantum average gate fidelity", which is derived from the quantum channel fidelity between two quantum processes. Since quantum gates are unitary operations, the average gate fidelity, and higher order moments, take a simple form and can be calculated analytically~\cite{Nie02,EAZ,MBKE}.

Similarly, \emph{partial} estimation of measurement errors by more efficient methods than full measurement tomography would be extremely valuable, but have yet to be considered in significant detail. There are a variety of questions to consider in this endeavor, such as how to deal with abundance of different measurement schemes across the various implementations, and the ambiguity regarding what the single parameter characterizing the error should be (see Ref.~\cite{OC} for two proposals of such distance measures). Since measurements are non-unitary processes, the quantum channel fidelity between the measurement and noise process is a very complex quantity and cannot be calculated analytically as is the case for quantum gates.

One of the main goals of this paper is to initiate research into scalable verification and characterization of measurement devices. Here, we discuss methods for characterizing the error of a quantum measurement and present protocols for estimating the performance of a measurement with respect to these error measures. The protocols are scalable in the number of qubits comprising the system. In addition, the protocols are very general and should be straightforward to implement in most systems. They require the ability to prepare a random state from the basis that constitutes the ideal measurement, and perform the noisy measurement (perhaps consecutively). As quantum systems continue to scale to larger sizes, and full tomography of the measurement becomes infeasible, the protocols provided here can be used as fast and straightforward methods to verify that the performance of a measuring device achieves some desired threshold.

Specifically, we propose the average measurement fidelity and error rate of a quantum measurement as natural measures of the error, and provide scalable, experimentally implementable protocols for estimating these quantities. The output of the protocol is a \emph{lower (upper) bound} on the average fidelity (error rate) of the measurement. Computing these bounds from the measurement data is straightforward, and generic estimates of how many trials need to be performed can be obtained using statistical methods. We provide a direct comparison of the time-complexity between our protocol and that of completely reconstructing the noisy measurement POVM elements via tomography. Our method scales independently of $d$ (where $d$ is the dimension of the system, e.g. $d=2^n$ for an $n$-qubit system) while a complete reconstruction scales as $O(d^3)$. Thus, while our method gives much less information than full tomography, it is scalable, straightforward to implement, and involves less post-processing of measurement data. Ideally, the protocols we provide here will be useful as a starting point for further analysis and research into more efficient tomographic methods of quantum measurements.

We focus mainly on the case where we attempt to implement a finite-dimensional projective measurement (also known as a projective-valued measure (PVM)), but instead implement a general quantum measurement in the form of a positive operator-valued measure (POVM). In particular, we analyze the following two scenarios:
\begin{enumerate}
\item The ideal measurement is a rank-1 PVM and we are only concerned with characterizing measurement probabilities.
\item The ideal measurement is a rank-1 PVM and we are concerned with simultaneously characterizing measurement probabilities and output states.
\end{enumerate}
Rank-1 PVM's are significant in QIP since each computational model mentioned above can achieve universal quantum computation using rank-1 PVM's. We leave extending our results to arbitrary-rank PVM's as a direction of future research, however we anticipate the results presented here can be carried over to these measurements.
Moreover, while the most general measurement model is a POVM, any POVM can be realized as a PVM on an extended Hilbert space by attaching ancilla systems to the Hilbert space of interest~\cite{Paul}. Thus, characterizing the quality of PVM's can potentially provide direct information about one's ability to implement POVM's.

The structure of the presentation is as follows. First, in Sec.~(\ref{sec:measurement}) we set notation, discuss the framework of PVM's and POVM's, and define the noisy measurement models. Next, in Sec.(\ref{sec:fidelity}), we define the \emph{quantum average measurement fidelity} and \emph{quantum average measurement error}, which are the operationally relevant metrics we use to compare measurement operations. In Sec.~(\ref{sec:protocol}) we present the main result for the first of the two scenarios listed above. We provide an expression for a lower (upper) bound on the fidelity (error rate), and present an experimental protocol for obtaining this bound.  In Sec.~(\ref{sec:derivation}) we give a derivation of the bound and show it satisfies certain necessary conditions to be useful in practice.  Sec.~(\ref{sec:lowerbound}) provides a discussion of the general validity of the bound and numerical examples for the case of a single-qubit system are analyzed. Our derived bounds are shown to be valid for \emph{every} example analyzed, confirming the discussion regarding the validity of the method. We discuss the extension of the method to rank-1 PVM's with output states in Sec.~(\ref{sec:outputstates}) and provide the experimental protocol for this case in Sec.~(\ref{sec:protocolstates}). The resource analysis for the protocols are provided in Sec.~(\ref{sec:resources}) and concluding remarks are made in Sec.~(\ref{sec:discussion}). The reader interested in the bounds and protocols without most of the technical details is referred to the text surrounding Eq.~(\ref{eq:lowerbound}) for rank-1 PVM's and the text surrounding Eq.~(\ref{eq:lowerboundstate}) for rank-1 PVM's with output states.

By definition, the average fidelity and error rate of a quantum measurement are equivalent (see Sec.~\ref{sec:fidelity}). Hence, the majority of the presentation is in terms of the average fidelity, however it is important to keep in mind that analogous results hold for the error rate.



\section{Measurement of quantum systems and the Error Model}\label{sec:measurement}

Denote the quantum system by $\mathcal{S}$ and suppose it is represented by a Hilbert space $\Hilb$ of dimension $d<\infty$ ($d=2^n$ for an $n$-qubit system). The most general type of measurement one can make is a ``positive operator-valued measure" (POVM). A POVM consists of a set $\{E_k\}$ of linear operators on $\Hilb$ that satisfy
\begin{eqnarray}
E_k &\geq& 0, \nonumber \\
\sum_kE_k &=& \Id. \label{eq:POVMconds}
\end{eqnarray}
If the input state to the measurement is $\sigma$, the probability of obtaining outcome ``$k$" is given by 
\begin{equation}
r_k = \tr\left(E_k \sigma\right).
\end{equation}
An important subset of POVM measurements are ``projection-valued measures" (PVM's), which correspond to the case of each $E_k$ being equal to a projection operator $\Pi_k$. Hence, in addition to the conditions listed in Eq.~(\ref{eq:POVMconds}), $\Pi_k^2=\Pi_k$ for each $k$. 

PVM's are a very general class of quantum measurements and, as previously mentioned, are sufficient for performing universal quantum computation in various computational models~\cite{Deutsch85,RB01}. For instance, computational basis measurements allow for universality in the standard circuit model and single-qubit projective measurements provide universality in MBQC. Other important examples of PVM's in quantum information theory are parity measurements, which are used extensively in quantum error-correction and fault tolerance~\cite{Got97,Got99,RDN,DS12}. It is important to note that, by Naimark's theorem~\cite{Paul}, any POVM can be implemented via a PVM on a larger Hilbert space. Hence, methods for determining error rates of PVM's can also give direct information regarding the error associated with performing general POVM measurements. 

PVM's are in a 1-1 correspondence with observables (Hermitian operators) $\mathcal{O}$, where $\mathcal{O}$ is non-degenerate if and only if the PVM consists of rank-1 projective elements . We write the observable $\mathcal{O}$ as
\begin{eqnarray}
\mathcal{O} &=& \sum_{k=1}^b \lambda_k \Pi_k,
\end{eqnarray}
where the eigenvalues $\lambda_k$ correspond to the measurement values and $\Pi_k:=\sum_{j=1}^{d_k}|\psi_j^k\rangle\langle \psi_j^k|$ is the rank-$d_k$ projector onto the eigenspace spanned by the eigenvectors $|\psi_j^k\rangle$.  Hence, $\sum_{k=1}^b d_k = d$ and the PVM is rank-1 if and only if $d_k = 1$ for every $k$. If $\mathcal{M}$ represents the ideal PVM without post-selection then $\mathcal{M}$ has the following action on each input state $\sigma$,
\begin{eqnarray}
\mathcal{M}(\sigma)&=&\sum_{k=1}^b \Pi_k\sigma\Pi_k = \sum_{k=1}^bp_k\left[\frac{\Pi_k\sigma \Pi_k}{\tr\left(\Pi_k\sigma\right)}\right],\label{eq:Measurement} 
\end{eqnarray}
where we have defined
\begin{gather}
p_k=p_k(\sigma)=\tr\left(\Pi_k\sigma\right).
\end{gather}

In the case of analyzing only measurement probabilities, the PVM is a mapping from the set of quantum states to the probability vectors $(p_1,...,p_b)$,
\begin{gather}
\mathcal{M}(\sigma) = (p_1(\sigma),...,p_b(\sigma))=(\tr\left(\Pi_1\sigma\right),...,\tr\left(\Pi_b\sigma\right)).\label{eq:pvec}
\end{gather}
This case is especially interesting since in many scenarios one is mainly interested in the output \emph{value} of the measurement rather than the output state itself. For instance, the output of a quantum algorithm is usually the value of a measurement in the computational basis. Hence, as long as the measurement values are obtained with correct probabilities, the measurement is deemed successful. 

In addition, in many current implementations of measurements, the action of performing the measurement operation to obtain the output value destroys or drastically alters the state of the system. For instance in photo-detection, which forms the basis of various measurement schemes in optical, atomic, and superconducting systems, the measurement consists of recording the occurrence of a photon. If the state of the system is encoded into a degree of freedom (mode) of the photon, such as polarization or frequency, detection of the photon can record the output of a measurement of this degree of freedom. In common photodetectors, such as avalanche photodiodes, the photon is 
lost in the process of creating a current through the photoelectric effect. Hence, the system encoding the information to be measured is destroyed, but the measurement output can be accessed.

An example where the system is not destroyed yet the state is not preserved from the measurement is the Nitrogen-Vacancy (NV) center in diamond~\cite{WJ06}. In this case, the processes of measurement and ground-state polarization are identical. Hence, while one can obtain the relevant measurement outcomes and statistics, the output state of the measurement is always the ground state. There are situations where preserving the measurement output state is useful. For instance, in any paradigm where quantum information is evolved by performing measurements, such as in measurement-based quantum computation~\cite{RB01}, one requires the state of the system to be well-preserved under measurements. In measurement-based quantum computing, quantum information is evolved by performing single-qubit measurements on a highly entangled initial state. Any degradation of the state or system caused by the measurement process can be highly detrimental to obtaining a high fidelity output.

%

In the general case of Eq.~(\ref{eq:Measurement}), the noisy measurement $\Eop$ is modeled by
\begin{gather}
\mathcal{E}(\sigma) = \sum_{k=1}^br_k(\sigma)\rho_k(\sigma),\label{eq:noisymeasurement}
\end{gather}
where we allow both the noisy measurement probabilities $r_k$ and output states $\rho_k$  to be functions of $\sigma$. By Eq.~(\ref{eq:Measurement}), $r_k(\sigma)$ and $\rho_k(\sigma)$ are ideally given by $p_k(\sigma)=\tr(\Pi_k\sigma)$ and $\frac{\Pi_k\sigma \Pi_k}{\tr\left(\Pi_k\sigma\right)}$ respectively. If we analyze only measurement probabilities, we have
\begin{gather}
\mathcal{E}(\sigma) = (r_1(\sigma),...,r_b(\sigma))=(\tr\left(E_1\sigma\right),...,\tr\left(E_b\sigma\right)),\label{eq:rvec}
\end{gather}
where $\Eop$ is allowed to be of a completely general form by assuming it is modeled by a POVM $\{E_k\}_{k=1}^b$. The first scenario we are interested in is to compare the probability distributions in Eq.~(\ref{eq:pvec}) and~(\ref{eq:rvec}). We analyze the more general case of measurement probabilities and output states (ie. comparing Eq.'s~(\ref{eq:Measurement}) and~(\ref{eq:noisymeasurement})) in Sec.~(\ref{sec:outputstates}). We first discuss the figure of merits we will use to compare the ideal measurement process $\mathcal{M}$ and noisy measurement process $\Eop$.

\section{Quantum Average Measurement Fidelity and Error}\label{sec:fidelity}

A completely question is, how should we compare the ideal and actual measurements $\Meas$ and $\Eop$? A set of criteria that a distance measure, $\Delta$, for comparing ideal and real quantum processes should satisfy has been given previously~\cite{GLN}. Currently, no known $\Delta$ satisfies all of these criteria simultaneously. Thus, one must settle for $\Delta$ to satisfy a subset of these criteria, in addition to other criteria that may be useful for the particular task at hand. 

As mentioned previously, the average gate fidelity $\overline{\gatefil}$ is a useful method for comparing an intended unitary operation $\mathcal{U}$ and actual quantum process $\Eop$. There are various reasons for the utility of the average gate fidelity, for instance, it satisfies the following properties:
\begin{enumerate}
\item There is a straightforward method for evaluating $\overline{\gatefil}$ (given a description of $\mathcal{U}$ and $\Eop$),
\item $\overline{\gatefil}$ has a well-motivated physical interpretation,
\item All states are taken into account in an unbiased manner when calculating $\overline{\gatefil}$,
\item $\overline{\gatefil}$ is experimentally accessible via efficient protocols.
\end{enumerate}
An important drawback of the average gate fidelity is that it is not a metric. We would like similar properties to hold for our method of comparing ideal and real measurements. Let us briefly outline how the average gate fidelity is derived from more general quantities, which will provide intuition for how to define our method for comparing measurements.

The average gate fidelity is derived from the state-dependent quantum channel fidelity, which is a standard method for comparing quantum operations. If $\Eop_1$ and $\Eop_2$ are quantum operations and $\sigma$ is a quantum state, the quantum channel fidelity between $\Eop_1$ and $\Eop_2$, denoted $F_{\Eop_1,\Eop_2}$, is given by the standard state fidelity between $\Eop_1(\sigma)$ and $\Eop_2(\sigma)$,
\begin{eqnarray}
F_{\Eop_1,\Eop_2}(\sigma)&=&F(\Eop_1(\sigma),\Eop_2(\sigma))\nonumber \\
&=&\left(\tr \sqrt{\sqrt{\Eop_1(\sigma)}\mathcal{E}_2(\sigma)\sqrt{\Eop_1(\sigma)}}\right)^2.\label{eq:channelfidelity}
\end{eqnarray}
When one of the operations in Eq.~(\ref{eq:channelfidelity}) is unitary (say $\Eop_2=\mathcal{U}$), the channel fidelity is called the quantum gate fidelity, and when $\sigma$ is pure ($\sigma=\kett\braa$), the gate fidelity takes the extremely simple form
\begin{eqnarray}
F_{\Eop_1,\mathcal{U}}(\kett\braa)&=&\tr\left(\Eop_1(\kett\braa)\: \mathcal{U}(\kett\braa)\right) \nonumber \\
&=& \braa \Lambda(\kett\braa)\kett\label{eq:Lambdafid}
\end{eqnarray}
where
\begin{equation}
\Lambda=\mathcal{U}^{\dagger}\circ \Eop.
\end{equation}
The \emph{average quantum gate fidelity}, denoted $\overline{\mathcal{F}_{\Eop_1,\mathcal{U}}}$, is obtained by integrating over all pure input states. The integral is taken over the unitarily invariant Haar measure (also known as the Fubini-Study measure) on the set of pure states~\cite{BZ}. In this paper we denote the Fubini-Study measure by $\mu$. This gives
\begin{eqnarray}
\overline{\mathcal{F}_{\Eop_1,\mathcal{U}}} &=& \int \tr\left(\Eop_1(\kett\braa)\: \mathcal{U}(\kett\braa)\right) d\psi \nonumber \\
&=& \int \braa \Lambda(\kett\braa)\kett d\psi \nonumber \\
&=& \frac{\sum_j \tr(A_j)\tr(A_j) + d}{d^2+d}\label{eq:avgkraus}
\end{eqnarray}
where $\{A_j\}$ is any set of Kraus operators for $\Lambda$~\cite{Nie02}. Thus, the average gate fidelity reduces to an extremely simple form because one of the operations is unitary.

Following this intuition, from Eq.'s~(\ref{eq:Measurement}),~(\ref{eq:noisymeasurement}), and~(\ref{eq:channelfidelity}), we have that for the ideal ($\mathcal{M}$) and noisy ($\Eop$) measurements
\begin{gather}
F(\Eop(\sigma),\Meas(\sigma)) \nonumber \\
= \left(\tr \sqrt{\sqrt{\sum_{k=1}^br_k\rho_k}\left(\sum_{k=1}^bp_k\left[\frac{\Pi_k\sigma \Pi_k}{\tr\left(\Pi_k\sigma\right)}\right]\right)\sqrt{\sum_{k=1}^br_k\rho_k}}\right)^2,\label{eq:measfid}
\end{gather}
where the state-dependence in $p_k$, $r_k$, $\rho_k$ is omitted for notational convenience. The operational significance of this quantity comes from the fact that it is just the standard fidelity between the two quantum states $\Eop(\sigma)$ and $\Meas(\sigma)$. More precisely, Eq.~(\ref{eq:measfid}) is related to the maximum distinguishability between probability distributions one could obtain using any POVM measurement on the states $\Eop(\sigma)$ and $\Meas(\sigma)$~\cite{FvdG}. 

Unlike the unitary case in Eq.~(\ref{eq:Lambdafid}), such an expression does not reduce to a simple form in general. However, ideally, the figure of merit we use to distinguish $\Meas$ and $\Eop$ will contain evenly weighted information from \emph{all} possible input states and also have direct operational significance. As a result, we define the \emph{quantum average measurement fidelity}, denoted $\FEM$, to be the Haar integral of Eq.~(\ref{eq:measfid}) over pure input states
\begin{equation}
\FEM = \int  F(\Eop(\kett),\Meas(\braa))d\psi.
\end{equation}
We also define
\begin{equation}
\rEM=1-\FEM\label{eq:rdef}
\end{equation}
to be the \emph{quantum average measurement error}. In the case of only analyzing measurement probabilities, that is $\mathcal{M}$ and $\Eop$ are given by Eq.'s~(\ref{eq:pvec}) and~(\ref{eq:rvec}) respectively, the states associated to $p_k$ and $r_k$ can be taken to be $\Pi_k$, which gives
\begin{align}
\FEM &= \int \left(\sum_{k=1}^b \sqrt{p_kr_k}\right)^2 d\psi \nonumber \\
&=\int\left( \sum_{k=1}^b \sqrt{\tr(\Pi_k\kett\braa)\tr(E_k\kett\braa)} \right)^2d\psi,\label{eq:fidmeasprobonly}
\end{align}
\begin{align}
\rEM&=1-\int\left( \sum_{k=1}^b \sqrt{\tr(\Pi_k\kett\braa)\tr(E_k\kett\braa)} \right)^2d\psi.\label{eq:rmeasprobs}
\end{align}

The direct relationship between $\rEM$ and $\FEM$ implies statements and bounds proven about one directly applies to the other. Thus, as mentioned in the introduction, we will phrase the majority of the discussion in terms of $\FEM$. Since $\FEM$ does not have a simple form like the average gate fidelity in Eq.~(\ref{eq:avgkraus}), our goal is to provide efficient methods for estimating $\FEM$ that can be experimentally implemented in a simple manner. 

An important reason for using the average measurement fidelity is that it provides a state-independent distance measure which can be connected to the diamond norm distance~\cite{Kit97} between quantum operations~\cite{BK11}. 
%
Computing the diamond norm is an exponentially hard task since one needs a complete description of the quantum operations. Thus, the diamond norm is neither straightforward to calculate nor experimentally accessible. However, it is commonly used in fault-tolerant analyses of threshold error rates of physical operations. Thus, information about the diamond norm provided by the average measurement error defined above can potentially provide information regarding the ability to perform fault-tolerant computation.
\section{Experimental Protocol: Rank-1 PVM's}\label{sec:protocol}
In this section, we look at obtaining a lower bound on $\FEM$ for measurement probabilities of rank-1 PVM's. Again, we note that upper bounds on $\rEM$ are equivalent to lower bounds on $\FEM$ and so, for both clarity and consistency, we phrase the following discussion in terms of only $\FEM$.

 There are $d$ PVM elements, $\{\Pi_1,...,\Pi_d\}$, each of which is a rank-1 projection operator. From Eq.~(\ref{eq:rvec}), the outcome of the measurement of a state $\kett\braa$ can be associated to a member from $\{1,...,d\}$ with frequency distribution 
\begin{gather}
(r_1,...,r_d) = (\tr(E_1 \kett\braa),...,\tr(E_d\kett\braa)),
\end{gather} 
where the $E_j$ represent the noisy POVM elements. 

Our expression for the lower bound is given by
\begin{align}
lb &= \frac{1+ d \overline{X}}{1+d},\label{eq:lowerbound}
\end{align}
where
\begin{align}
\overline{X} &= \frac{1}{d^2}\displaystyle{\sum_{(l,m) \in \mathcal{D}}}\sqrt{u_lu_m},\\ \label{eq:Xbardef}
u_l &= \tr(\Pi_lE_l),
\end{align}
and
\begin{equation}
\mathcal{D}=\{0,...,d-1\} \times \{0,...,d-1\}.
\end{equation}
The $u_l$ measure the overlap between the $l$'th ideal PVM element and $l$'th noisy POVM element. Thus, $lb$ can intuitively be thought of as a parameter that measures how well ideal PVM elements are preserved when input to the noisy measurements (with the inclusion of dimensional factors).


\bigskip

 \underline{Goal}: Obtain a lower bound, $lb$, for $\FEM$ as defined in Eq.~(\ref{eq:rmeasprobs}). 

\bigskip

\noindent The experimental protocol to obtain $lb$ is as follows:

\bigskip

 \underline{Protocol}:

\medskip


\medskip

\noindent Step 1: Choose a pair of indices $(l,m)$ uniformly at random from $\mathcal{D}$.

\bigskip

\noindent Step 2: Independently, for both $j = l$ and $j=m$,

\bigskip

a): Prepare the quantum state $|\psi_j\rangle$, perform the noisy measurement $\Eop$ on $\Pi_j=|\psi_j\rangle\langle \psi_j|$, and record whether outcome ``$j$" is obtained,


\bigskip

b): Repeat a) many times and denote the frequency of obtaining ``$j$" by $\hat{u}_j$, that is, $\hat{u}_j$ is an estimator of $u_j=\tr(\Pi_jE_j)$,

\medskip

\noindent (see Sec.~\ref{sec:trials1} for a discussion of the number of repetitions required to estimate $u_j$ to a desired accuracy and confidence).
\bigskip

\noindent Step 3: Repeat Steps 1 and 2 $K$ times, where $K$ is dictated by the desired accuracy and confidence in estimating $lb$

\medskip

\noindent (see Sec.~\ref{sec:trials2} for a discussion of the size of $K$).


\bigskip

\noindent Step 4: Compute an estimator $\hat{lb}$ for the lower bound $lb$ of $\FEM$ defined in Eq.~(\ref{eq:lowerbound}) via the formula


\begin{eqnarray}
\hat{lb} &=& \frac{1+ d\hat{\overline{X}}}{1+d},
\end{eqnarray}
where

\begin{align}
\hat{\overline{X}} &=  \frac{1}{K}\displaystyle{\sum_{(k_1,k_2)}\sqrt{\hat{u}_{k_1}\hat{u}_{k_2}}} \label{eq:Xbardef}
\end{align}
is an estimator of $X$ defined in Eq.~(\ref{eq:Xbardef}) and the $\{(1_1,1_2),...,(K_1,K_2)\}$ are the $K$ trials dictated by Step 3. 

\medskip

This concludes the protocol.

\medskip

There are various important points about the protocol that should be emphasized. First, the number of trials required in Steps 2b) and 3 are independent of $d$. Thus, the time-complexity of the entire protocol is independent of $d$, and depends only on the desired accuracy and confidence of the estimate $\hat{lb}$ of $lb$ (see Sec.~\ref{sec:resources}). Second, $lb$ can be estimated from the above protocol using only:
\begin{enumerate}
\item Applications of the noisy measurement and
\item The ability to prepare the $d$ pure input states $|\psi_j\rangle$. 
\end{enumerate}
Lastly, it is straightforward to show the following two properties of $lb$ hold: (see Sec.~\ref{sec:necessary})
\begin{enumerate}
\item In the limit of $\FEM \uparrow 1$, 
\begin{equation}
lb \uparrow 1.
\end{equation}
\item $lb$ scales well in $d$, that is, if each $u_k=\tr(\Pi_kE_k)$ is on the order of $1-\delta$ then, as $d \rightarrow \infty$, $lb \rightarrow 1-\delta$ (and does not converge to 0 or some other small constant).
\end{enumerate}
These are clearly necessary conditions for $lb$ to be a good lower bound on $\FEM$. Property 1 implies that, loosely speaking, in the small error limit $lb$ can be taken as an estimate of $\FEM$. Property 2 implies that as $d \rightarrow \infty$, the lower bound remains on the order of $1-\delta$. This is important because if the lower bound generically converges to 0 (or some other constant) as $d\rightarrow \infty$, then the lower bound is ineffective at providing useful information about the noise.

\section{Derivation of the Lower Bound}\label{sec:derivation}
We need to show
\begin{eqnarray}
lb &:=& \frac{1+\overline{X}d}{1+d}\nonumber \\
&=&  \frac{d+ \sum_{k=1}^du_k + \sum_{l\neq m}\sqrt{u_lu_m}}{d(d+1)}\label{eq:lowerbound}
\end{eqnarray}
is a lower bound for $\FEM$. 
First, we have from Eq.~(\ref{eq:fidmeasprobonly}), 
\begin{gather}
F(\Eop(\sigma),\Meas(\sigma)) = \left(\sum_k\sqrt{r_kp_k}\right)^2, \nonumber \\
\end{gather}
where the state-dependence is implicit in the $p_k$ and $r_k$. Taking the integral over all pure states gives
\begin{gather}
\int F(\Eop(\kett\braa),\Meas(\kett\braa))d\psi \nonumber \\
= \sum_{k=1}^d\left[\int r_kp_kd\psi\right] + \sum_{l\neq m}\left[\int\sqrt{r_lr_mp_lp_m}d\psi\right].\label{eq:twosums}
\end{gather}
With these tools in hand, let us look at the sums in Eq.~(\ref{eq:twosums}) separately. 
\subsection{$\displaystyle \sum_{k=1}^d\int r_kp_k d\psi$}\label{sec:firstset}
We have 
\begin{align}
\int r_kp_k d\psi &= \int \tr(E_k\kett\braa) \tr(\Pi_k\kett\braa)d\psi \nonumber \\
&= \int \tr\left(\left[E_k\otimes \Pi_k\right] \kett\braa\otimes \kett\braa\right)d\psi\nonumber \\
&= \tr\left(\left[E_k\otimes \Pi_k\right] \int \kett\braa^{\otimes 2} d\psi\right).
\end{align}
To compute this integral we use Schur's Lemma which states that for any positive integer $t$,~\cite{RBSC,MBKE}
\begin{equation}
\int |\psi\rangle\langle \psi | ^{\otimes t} d\psi = \frac{\Pi_{\text{sym}}(t,d)}{\tr \left[\Pi_{\text{sym}}(t,d)\right]},\label{eq:sym}
\end{equation}
where $\Pi_{\text{sym}}(t,d)$ is the projector onto the symmetric subspace of the $t$-partite Hilbert space (whose $t$ factor spaces each have dimension $d$). The symmetric subspace of a $t$-partite Hilbert space consists of states that are left unchanged under permutations of the factor spaces. Reasoning for why Eq.~(\ref{eq:sym}) holds is as follows. First, by symmetry of $|\psi\rangle\langle \psi | ^{\otimes t}$, $\int |\psi\rangle\langle \psi | ^{\otimes t} d\psi$ must only have support on the symmetric subspace. In addition, since this operator is invariant under multiplication by any unitary operator of the form $U^{\otimes t}$, and has unit trace, Schur's Lemma implies that it must be equal to the normalized projector onto the symmetric subspace, which is just the right hand side of Eq.~(\ref{eq:sym}).
This implies
\begin{align}
\int r_kp_k d\psi &= \frac{2\tr\left(\left[E_k \otimes \Pi_k\right] \Pi_{\text{sym}}(2,d) \right)}{d(d+1)}.\label{eq:rkpk}
\end{align}

Now, one can show that $\Pi_{\text{sym}}(t,d)$ is equal to the normalized sum of the $t!$ elements in the group of permutation operators on the $t$-partite Hilbert space
\begin{align}
\Pi_{\text{sym}}(t,d)&=\frac{1}{t!}\sum_\sigma P_\sigma.
\end{align}
This can be shown by first noting that the set of all permutations is a subgroup of the unitary group. Thus the square of the normalized sum is just equal to the normalized sum itself
\begin{align}
\left(\frac{1}{t!}\sum_\sigma P_\sigma\right)^2&=\frac{1}{t!}\sum_\sigma P_\sigma.
\end{align}
Hence, since $\frac{1}{t!}\sum_\sigma P_\sigma$ commutes with any permutation $P_\tau$, it must be equal to the projection operator onto the symmetric subspace. As a simple example, in the case of $t=2$, there are two permutation operators, $\mathbbm{1}$, and the SWAP operation which swaps the two factor spaces. Thus
\begin{align}
\Pi_{\text{sym}}(2,d) &=\frac{\Id\otimes \Id + \text{SWAP}}{2}.\label{eq:twod}
\end{align}

Eq.'s~(\ref{eq:rkpk}) and (\ref{eq:twod}) imply
\begin{align}
\int r_kp_k d\psi&=\frac{\tr\left(\left[E_k \otimes \Pi_k\right] \left[\Id\otimes \Id + \text{SWAP}\right] \right)}{d(d+1)} \nonumber \\
&=\frac{\tr(E_k \Pi_k) + \tr(E_k)\tr(\Pi_k)}{d(d+1)} \nonumber \\
&=\frac{\tr(E_k \Pi_k) + d\tr(E_k\frac{\Id}{d})}{d(d+1)}, \nonumber \\
\end{align}
and so
\begin{align}
\sum_{k=1}^d\int r_kp_k d\psi &= \sum_{k=1}^d \frac{\tr(E_k \Pi_k) + d\tr(E_k\frac{\Id}{d})}{d(d+1)} \nonumber \\
&= \displaystyle{\frac{\sum_{k=1}^du_k + d}{d(d+1)}},\label{eq:exactfirstset}
\end{align}
since $\sum_{k=1}^dE_k = \Id$. This gives the first two terms in the numerator in Eq.~(\ref{eq:lowerbound}).
\subsection{$\displaystyle\sum_{l\neq m}\int\sqrt{r_lr_mp_lp_m}d\psi$}\label{sec:secondset}
We have
\begin{widetext}
\begin{gather}
\int\sqrt{r_lr_mp_lp_m}d\psi = \int \sqrt{\tr(\Pi_l\kett\braa) \tr(E_l\kett\braa)}\sqrt{\tr(\Pi_m\kett\braa) \tr(E_m\kett\braa) }d\psi.
\end{gather}
Computing the above integral analytically is difficult because of the square root in the argument.
Now we assume the following inequality holds (see Sec.~\ref{sec:lowerbound})
\begin{gather}
\int \sqrt{\tr(\Pi_l\kett\braa) \tr(E_l\kett\braa)}\sqrt{\tr(\Pi_m\kett\braa) \tr(E_m\kett\braa) }d\psi \nonumber \\
\geq \int \sqrt{\tr(\Pi_l\kett\braa) \tr(\Pi_lE_l\Pi_l(\kett\braa))} \sqrt{\tr(\Pi_m\kett\braa)\tr(\Pi_mE_m\Pi_m(\kett\braa))}d\psi \nonumber \\
= \int \sqrt{\langle\psi_l|E_l|\psi_l\rangle}\tr(\Pi_l(\kett\braa)) \sqrt{\langle\psi_m|E_m|\psi_m\rangle}\tr(\Pi_m(\kett\braa)) d\psi \nonumber \\
= \sqrt{u_lu_m}\int \tr(\Pi_l(\kett\braa)) \tr(\Pi_m(\kett\braa)) d\psi.\label{eq:Qu}
\end{gather}
\end{widetext}
Note that for each $j$ it is generally not true that $E_j-u_j\Pi_j$ is positive semidefinite (if this were the case then the above inequality would always hold).

From Eq.~(\ref{eq:sym}) we have since $\Pi_l$ and $\Pi_m$ are projectors onto orthogonal subspaces,
\begin{gather}
\int \tr(\Pi_l(\kett\braa)) \tr(\Pi_m(\kett\braa)) d\psi \nonumber \\
= \int \tr\left[\left(\Pi_l\otimes \Pi_m \right) \kett\braa \otimes \kett\braa\right]d\psi \nonumber \\
= \frac{\tr\left( \Pi_l  \Pi_m \right) + \tr\left(\Pi_l \right) \tr\left( \Pi_m \right)}{d(d+1)}\nonumber \\
=\frac{1}{d(d+1)}.
\end{gather}
Thus
\begin{eqnarray}
\int\sqrt{r_lr_mp_lp_m}d\psi &\geq& \frac{ \sqrt{\tr\left(E_l\Pi_l \right) \tr\left( E_m \Pi_m \right)}}{d(d+1)} \nonumber \\
&=& \frac{\sqrt{u_l u_m}}{d(d+1)},\label{eq:secondsetlb}
\end{eqnarray}
which gives the last term in the numerator of Eq.~(\ref{eq:lowerbound}).

In total, assuming the inequality in Eq.~(\ref{eq:Qu}) holds, we have that
\begin{align}
\frac{1+d\overline{X}}{1+d}=\frac{d+ \sum_{k=1}^du_k + \sum_{l\neq m}\sqrt{u_lu_m}}{d(d+1)}
\end{align}
is a lower bound for $\FEM$. Before analyzing the validity of the lower bound, we briefly compare its expression with that of the average gate fidelity and show that it satisfies certain necessary conditions to be useful in practice.

\subsection{Comparison With Quantum Gate Fidelity}\label{sec:comparison}

There is a nice parallel between the lower bound on the average measurement fidelity and the exact expression for the average quantum gate fidelity. As previously mentioned, the quantum average gate fidelity takes a particularly nice form because one of the operations is unitary~\cite{Nie02}. Indeed, if one compares the unitary $\mathcal{U}$ and quantum operation $\Eop$, the average gate fidelity between $\mathcal{U}$ and $\Eop$, $\overline{\mathcal{F}_{\Eop,\mathcal{U}}}$, is given by
\begin{align}
\overline{\gatefil} = \frac{\overline{A}d+1}{d+1},
\end{align}
where
\begin{align}
\overline{A}=\frac{1}{d^2}\sum_k\tr(A_k)\tr(A_k^{\dagger}),
\end{align}
and $\{A_k\}$ is any set of Kraus operators for the quantum operation $\Lambda=\mathcal{U}^{\dagger}\circ \Eop$.

Our expression for a lower bound on the average measurement fidelity $\FEM$ takes a similar form:
\begin{equation}
lb=\frac{\overline{X}d+1}{d+1}
\end{equation}
where
\begin{equation}
\overline{X}=\frac{1}{d^2}\sum_{l,m}\sqrt{\tr(\Pi_lE_l)\tr(\Pi_mE_m)}.
\end{equation}
In some sense this is not surprising since each quantity is a Haar integral over functions of two copies of a quantum state $\kett\braa$. A direction of further research is to understand properties of the average measurement fidelity in more detail, and draw more parallels with well-known distinguishability measures such as the average gate fidelity.

\subsection{Necessary Conditions For $lb$ To Be a Useful Lower Bound on $\FEM$}\label{sec:necessary}

In this subsection we discuss two necessary conditions $lb$ must satisfy in order to be a useful lower bound on $\FEM$;
%
\begin{enumerate}
\item Limit of No Error: As the measurement error goes to 0, $lb \uparrow 0$.
\item Scaling in $d$: $lb$ scales well in the dimension $d$ of the system.
\end{enumerate}
\medskip 
Here, we show $lb$ satisfies both of these criteria.

\medskip

\subsubsection{Limit of No Error}\label{sec:Limitnoerrors}

Suppose the measurement error goes to $0$ in that the POVM elements converge to the ideal PVM elements
\begin{eqnarray}
E_j \rightarrow \Pi_j.
\end{eqnarray}
This implies for each $j \in \{1,...,d\}$,
\begin{equation}
u_j = \tr(\Pi_jE_j) \uparrow 1.
\end{equation}
Hence, from Eq.'s~(\ref{eq:lowerbound}) and (\ref{eq:Xbardef}), 
\begin{eqnarray}
lb \uparrow 1.
\end{eqnarray}
Thus $lb$ satisfies the necessary condition of converging to $1$ in the limit of no errors.

\subsubsection{Scaling in $d$}\label{sec:scaling}

Let us now explicitly show that $lb$ scales well in $d$. More precisely, if each $u_k=\tr(\Pi_kE_k)$ is on the order of $1-\delta$ for some $\delta > 0$ then, as $d \rightarrow \infty$, $lb \rightarrow 1-\delta$. Hence, the lower bound does not generically converge to a constant value that is independent of the $E_k$. Such an effect would render the lower bound useless as the system size grows large.

Again, by Eq.'s~(\ref{eq:lowerbound}) and (\ref{eq:Xbardef}) we see that if each $u_k$ satisfies
\begin{align}
u_k&\sim1-\delta
\end{align}
for some $\delta > 0$ then
\begin{align}
lb&\sim\frac{1+\overline{X}d}{1+d}\nonumber \\
&=\frac{1+(1-\delta)d}{1+d}.
\end{align}
Hence as $d\rightarrow \infty$,
\begin{align}
lb \rightarrow 1-\delta,
\end{align}
which is what we wanted to show.

\section{Validity of the Lower Bound}\label{sec:lowerbound}

The lower bound in Eq.~(\ref{eq:lowerbound}) is valid provided the inequality in Eq.~(\ref{eq:Qu}) holds. The goal of this section is to show that this inequality holds in very general situations. To set notation, we define
\begin{widetext}
\begin{eqnarray}
f_{l,m}(\kett) &=& \sqrt{\tr(\Pi_l\kett\braa) \tr(E_l\kett\braa)} \sqrt{\tr(\Pi_m\kett\braa) \tr(E_m\kett\braa) },\nonumber \\
g_{l,m}(\kett) &=& \sqrt{\tr(\Pi_l\kett\braa) \tr(\Pi_lE_l\Pi_l\kett\braa)} \sqrt{\tr(\Pi_m\kett\braa) \tr(\Pi_mE_m\Pi_m\kett\braa) } \nonumber \\
&=& \sqrt{u_lu_m} \tr(\Pi_l\kett\braa) \tr(\Pi_m\kett\braa) \label{eq:fandgdef}
\end{eqnarray}
\end{widetext}
so that we want to show
\begin{gather}
\int [f_{l,m}(\kett) - g_{l,m}(\kett)]d\psi \geq 0. \label{eq:ineq}
\end{gather}
We demonstrate that Eq.~(\ref{eq:ineq}) should hold by showing that the set of states for which $f_{l,m}(\kett) - g_{l,m}(\kett) < 0$ is small and, even if $\kett$ satisfies $f_{l,m}(\kett) - g_{l,m}(\kett) < 0$, $\left|f_{l,m}(\kett) - g_{l,m}(\kett)\right|$ will be small.
So, let us look at the measure of states $\kett$ that could satisfy 
\begin{align}
f_{l,m}(\kett) - g_{l,m}(\kett) < 0
\end{align}
and what $\left|f_{l,m}(\kett) - g_{l,m}(\kett)\right|$ will look like for such a state. 

First note that
\begin{equation}
\int [f_{l,m}(\kett) - g_{l,m}(\kett)]d\psi = \int k_{l,m}(\kett)h_{l,m}(\kett) d\psi,
\end{equation}
where
\begin{align}
k_{l,m}(\kett)&:=\sqrt{\tr(\Pi_l\kett\braa) \tr(\Pi_m\kett\braa)},\\
h_{l,m}(\kett)&:=\sqrt{\tr(E_l\kett\braa) \tr(E_m\kett\braa)}\\
&\: \:\: \:-\sqrt{u_lu_m}\sqrt{\tr(\Pi_l\kett\braa) \tr(\Pi_m\kett\braa)}.
\end{align}
Since $k_{l,m}(\kett) \geq 0$ for all $\kett$, we have
\begin{align}
f_{l,m}(\kett)-g_{l,m}(\kett) < 0 \Leftrightarrow h_{l,m}(\kett) < 0.
\end{align}
However, if $h_{l,m}(\kett) < 0$ then either
\begin{equation}
\braa E_l-u_l\Pi_l\kett < 0 \: \: \text{or} \: \:  \braa E_m-u_m\Pi_m\kett < 0.\label{eq:condd1}
 \end{equation}
Without loss of generality, suppose $\braa E_l-u_l\Pi_l\kett < 0$. Thus, if $\kett$ satisfies $f_{l,m}(\kett) - g_{l,m}(\kett) < 0$ then $\braa E_l-u_l\Pi_l\kett < 0$.

As a matrix written in the $\{|\psi_k\rangle\}$ basis, we have $E_l-u_l\Pi_l$ is equal to $E_l$ except in the $(l,l)$'th entry which is 0. For instance if $l=0$ then $E_0-u_0\Pi_0$ is given by
\begin{center}
\begin{equation}
\begin{bmatrix}
0  & E_0^{0,1}     &   \hdots    &  \hdots      & E_0^{0,d-2}     & E_0^{0,d-1}  \\
E_0^{1,0}  & E_0^{1,1} & \ddots & \ddots &        & \vdots  \\
\vdots   & \ddots & \ddots & \ddots & \ddots &  \vdots  \\
 \vdots  & \ddots & \ddots & \ddots & \ddots & \vdots \\
E_0^{d-2,0}   &        & \ddots & \ddots & \ddots & \vdots  \\
E_0^{d-1,0}   & \hdots    &  \hdots      &  \hdots   &\hdots      & E_0^{d-1,d-1} 
\end{bmatrix}.
\end{equation}
\end{center}


\noindent Hence, $E_l-u_l\Pi_l$ is not positive semidefinite unless the off-diagonals in the $l$'th row (column) are equal to 0. This is because, since $E_l$ is positive semidefinite, the following bound on the off-diagonal elements holds~\cite{HJ}
\begin{equation}
\left|E_l^{i,j}\right| \leq \sqrt{E_l^{i,i}}\sqrt{E_l^{j,j}}.
\end{equation}
Now, note that if $\kett = |\psi_l\rangle$ then $\braa E_l-u_l\Pi_l \kett =0$, and if $\kett$ is orthogonal to $|\psi_l\rangle$, $\braa E_l-u_l\Pi_l\kett \geq 0$. Hence, there is some small region in the space of pure states centered around the state $|\psi_l\rangle$ for which one could have
\begin{equation}
\braa E_l-u_l\Pi_l \kett < 0.
\end{equation}
That is, the function on pure states defined by $\braa E_l-u_l\Pi_l \kett$ has a minimum (which can be less than 0) that is achieved for some state close to $|\psi_l\rangle$. As $\kett$ moves farther away from $|\psi_l\rangle$, and accumulates more amplitude in the subspace orthogonal to $|\psi_l\rangle$, we have
\begin{equation}
\braa E_l-u_l\Pi_l \kett \geq 0.
\end{equation}
Thus, Eq.~(\ref{eq:condd1}) can only be satisfied if $\kett$ has large enough amplitude in the subspace defined by $|\psi_l\rangle$. The amount of amplitude that is required depends on the size of the off-diagonal elements (coherence) in the $l$'th column (row) of $E_l$. If the coherence is not too large then $\kett$ will need to have large amplitude in $|\psi_l\rangle$ to satisfy Eq.~(\ref{eq:condd1}), and as the coherence grows larger, $\kett$ can potentially satisfy Eq.~(\ref{eq:condd1}) without being very close to $|\psi_l\rangle$. 

The key point is that the measure of the set of states for which $\kett$ is close to $|\psi_l\rangle$ (or $|\psi_m\rangle$) will be small (and by Levy's Lemma~\cite{Led01} decreases exponentially in the dimension $d$). Moreover, if $\kett$ has large amplitude in $|\psi_l\rangle$, and so $\braa E_l-u_l\Pi_l\kett < 0$ may occur, then $\tr(\Pi_m\kett\braa)$ is small since $|\psi_l\rangle$ and $|\psi_m\rangle$ are orthogonal. Thus, by definition, if $\kett$ is close to either $|\psi_l\rangle$ or $|\psi_m\rangle$, $k_{l,m}(\kett)$ will be small. Hence, $k_{l,m}(\kett)$ acts like a \emph{modulating factor} in the expression for $f_{l,m}(\kett)-g_{l,m}(\kett)$ to ensure $f_{l,m}(\kett)-g_{l,m}(\kett)$ is close to 0 when $\kett$ is close to either of $|\psi_l\rangle$ or $|\psi_m\rangle$. 

 In total then, the measure of states for which $\kett$ is close to $|\psi_l\rangle$ or $|\psi_m\rangle$ is small and, for any such state, $f_{l,m}(\kett)-g_{l,m}(\kett)$ will be close to 0. Equivalently, the set of states for which $f_{l,m}(\kett)-g_{l,m}(\kett) < 0$ is small and, for any such state, $\left|f_{l,m}(\kett)-g_{l,m}(\kett)\right|$ will be close to 0. This is what we wanted to show and so we expect Eq.~(\ref{eq:ineq}) to hold for most practical cases.

Note that the bounds are guaranteed to be valid if there is no coherence at all, that is, if the POVM elements are diagonal in the $\{|\psi_k\rangle\}$ basis. In the next section we obtain a sufficient condition for Eq.~(\ref{eq:ineq}) to hold in the case of a single qubit. In Sec.~\ref{sec:numerics} we perform a detailed numerical investigation of the single-qubit case and show our lower bound always holds, that is, for \emph{all} values of coherence magnitude, the bounds are valid. This provides evidence that the bounds will be valid in large generality.
Clearly, a more in-depth investigation of sufficient conditions for the bounds to be valid is desirable, however the above argument and numerical results of Sec.~(\ref{sec:numerics}) indicate that the bounds and protocol will be valid in most practical cases.

\subsection{Sufficient Condition For the Single-Qubit Case}

Using the notation of the previous section, for a single qubit, the lower bound in Eq.~(\ref{eq:lowerbound}) is valid if
\begin{gather}
\int\left[ f_{0,1}(\kett)- g_{0,1}(\kett)\right]d\psi \geq 0.\label{eq:ineq1}
\end{gather}
Our goal in this section is to obtain a sufficient condition for when Eq.~(\ref{eq:ineq1}) holds. 

We use the Bloch sphere representation of a single qubit:
\begin{equation}
\kett=\cos\left(\frac{\theta}{2}\right)|0\rangle + e^{\i\phi}\sin\left(\frac{\theta}{2}\right)|1\rangle
\end{equation}
where $\theta \in [0,\pi)$, $\phi\in [0,2\pi)$. Hence, the POVM  elements are given by
\begin{gather}
E_0 = \left(\begin{array}{cc}
u_0 & \gamma \\
\gamma & \tr(E_0\Pi_1)
\end{array} \right),
\end{gather}
\begin{gather}
E_1 = \left(\begin{array}{cc}
\tr(E_1\Pi_0) & -\gamma \\
-\gamma & u_1
\end{array} \right),
\end{gather}
where, since the results will depend on the magnitude of the coherence in the POVM elements, we assume without loss of generality that $\gamma \in \mathbbm{R}$ and $\gamma > 0$.

Since our goal is obtaining a lower bound on $\int \left[f_{0,1}(\kett)-g_{0,1}(\kett)\right]d\psi$, we need to understand when one of $\braa E_0-u_0\Pi_0 \kett < 0$ or $\braa E_1-u_1\Pi_1 \kett < 0$ holds since this is a necessary condition for $f_{0,1}(\kett)-g_{0,1}(\kett)< 0$ (see the discussion surrounding Eq.~(\ref{eq:condd1}) in the previous section). Without loss of generality, let us assume
\begin{equation}
\braa E_0-u_0\Pi_0 \kett < 0. \label{eq:weakcond}
\end{equation}
We have
\begin{align}
\braa E_0-u_0\Pi_0\kett &= 2\text{Re}\left(\gamma \cos\left(\frac{\theta}{2}\right)e^{i\phi}\sin\left(\frac{\theta}{2}\right)\right) \nonumber \\
&\:\:\: + \tr(E_0\Pi_1)\sin^2\left(\frac{\theta}{2}\right)
\end{align}
and it is straightforward to show $\braa E_0-u_0\Pi_0 \kett < 0$ is satisfied if
\begin{eqnarray}
\phi &\in& \left(\frac{\pi}{2},\frac{3\pi}{2}\right), \nonumber \\
\theta &\in& \left[0,2\:\text{arccot}\left(-\frac{\tr(E_0\Pi_1)}{2\gamma\cos(\phi)}\right)\right].\label{eq:phithetaconds}
\end{eqnarray} 
Hence, denoting the set of states that satisfy $\braa E_0-u_0\Pi_0 \kett < 0$ by $A_{0,\gamma}$,
we have that $A_{0,\gamma}$ is contained in the set of all states with $\phi$ and $\theta$ given by Eq.~(\ref{eq:phithetaconds}). As expected, the measure of $A_{0,\gamma}$, $\mu(A_{0,\gamma})$, is extremely small for weak coherence and grows larger as the coherence increases in magnitude. 

The uniform measure on the Bloch sphere has density function
\begin{equation}
\frac{1}{4\pi}\sin(\theta).
\end{equation}
Hence, from Eq.'s~(\ref{eq:phithetaconds}), $\mu(A_{0,\gamma})$ is at most
\begin{gather}
\frac{1}{4\pi}\int_{\frac{\pi}{2}}^{\frac{3\pi}{2}}\int_{0}^{2\text{arccot}\left(-\frac{\tr(E_0\Pi_1)}{2\gamma\cos(\phi)}\right)}\sin(\theta)d\theta d\phi \nonumber \\
= 
\frac{1}{4}-\frac{1}{4\pi}\int_{\frac{\pi}{2}}^{\frac{3\pi}{2}}\cos\left(2\:\text{arccot}\left(-\frac{\tr(E_0\Pi_1)}{2\gamma\cos(\phi)}\right)\right)d\phi.\label{eq:boundmeas}
\end{gather}
Now, if $\kett \in A_{0,\gamma}$
\begin{align}
\theta &< 2\: \text{arccot}\left(-\frac{\tr(E_0\Pi_1)}{2\gamma\cos(\phi)}\right),
\end{align}
and so
\begin{align}
\tr(\Pi_1\kett \braa) &\leq \sin^2\left(\text{arccot}\left(-\frac{\tr(E_0\Pi_1)}{2\gamma\cos(\phi)}\right)\right) \nonumber \\
&=\frac{\cos^2(\phi)}{\cos^2(\phi)+\left(\frac{\tr(E_0\Pi_1)}{2\gamma}\right)^2}.\label{eq:sin2arccot}
\end{align}
Thus, we have that for weak coherence $\gamma$, the set of states which satisfy Eq.~(\ref{eq:weakcond}) will have small measure bounded by Eq.~(\ref{eq:boundmeas}). Using a symmetric argument for the state $|1\rangle$ gives the following set of equations,
\begin{widetext}
\begin{align}
\mu(A_{0,\gamma}) &\leq  \frac{1}{4}-\frac{1}{4\pi}\int_{\frac{\pi}{2}}^{\frac{3\pi}{2}}\cos\left(2\text{arccot}\left(-\frac{\tr(E_0\Pi_1)}{2\gamma\cos(\phi)}\right)\right), \nonumber \\
\mu(A_{1,\gamma}) &\leq  \frac{1}{4}-\frac{1}{4\pi}\int_{\frac{\pi}{2}}^{\frac{3\pi}{2}}\cos\left(2\text{arccot}\left(-\frac{\tr(E_1\Pi_0)}{2\gamma\cos(\phi)}\right)\right), \nonumber \\
\text{If} \:\: \kett \in A_{0,\gamma} \;, \: \tr(\Pi_1\kett \braa) &\leq  \frac{\cos^2(\phi)}{\cos^2(\phi)+\left(\frac{\tr(E_0\Pi_1)}{2\gamma}\right)^2} \: \: \text{with}\:\: \phi\in \left(\frac{\pi}{2},\frac{3\pi}{2}\right),\nonumber \\
\text{If}\:\:  \kett \in A_{1,\gamma} \;, \: \tr(\Pi_0\kett \braa) &\leq  \frac{\cos^2(\phi)}{\cos^2(\phi)+\left(\frac{\tr(E_1\Pi_0)}{2\gamma}\right)^2} \: \: \text{with}\:\: \phi\in \left(\frac{\pi}{2},\frac{3\pi}{2}\right). \label{eq:relns}
\end{align}
 We can now obtain a sufficient condition for
\begin{gather}
\int \left[f_{0,1}(\kett)-g_{0,1}(\kett)\right] d\psi = \sum_{j=0}^1\int_{A_{j,\gamma}}  \left[f_{0,1}(\kett)-g_{0,1}(\kett)\right] d\psi  +  \int_{(A_{0,\gamma}\cup A_{1,\gamma})^c} \left[f_{0,1}(\kett)-g_{0,1}(\kett)\right]  d\psi \geq 0.
\end{gather}
\end{widetext}
By the set of equations in Eq.~(\ref{eq:relns}) one can show
\begin{gather}
\int_{A_{0,\gamma}} \left[f_{0,1}(\kett)-g_{0,1}(\kett)\right] d\psi  \nonumber \\
\geq \int_{A_{0,\gamma}} 0-\sqrt{u_0u_1}\tr(\Pi_0\kett\braa)\tr(\Pi_1\kett\braa)d\psi \nonumber \\
= -\frac{\sqrt{u_0u_1}}{2\pi}\int_{\frac{\pi}{2}}^{\frac{3\pi}{2}}\sin^4\left(\text{arccot}\left(-\frac{\tr(E_1\Pi_0)}{2\gamma\cos(\phi)}\right)\right) d\phi.
\end{gather}
By Eq.~(\ref{eq:sin2arccot})
\begin{gather}
\int_{A_{0,\gamma}} \left[f_{0,1}(\kett)-g_{0,1}(\kett)\right] d\psi\nonumber \\
\geq -\frac{\sqrt{u_0u_1}}{2\pi}\int_{\frac{\pi}{2}}^{\frac{3\pi}{2}}\left[\frac{\cos^2(\phi)}{\cos^2(\phi)+\left(\frac{\tr(E_0\Pi_1)}{2\gamma}\right)^2}\right]^2 d\phi,\nonumber \\
=: -\delta_{0,\gamma}
\end{gather}
and similarly
\begin{gather}
\int_{A_{1,\gamma}} \left[f_{0,1}(\kett)-g_{0,1}(\kett)\right] d\psi \nonumber \\ 
\geq-\frac{\sqrt{u_0u_1}}{2\pi}\int_{\frac{\pi}{2}}^{\frac{3\pi}{2}}\left[\frac{\cos^2(\phi)}{\cos^2(\phi)+\left(\frac{\tr(E_1\Pi_0)}{2\gamma}\right)^2}\right]^2 d\phi\nonumber \\
=: -\delta_{1,\gamma}.
\end{gather}
Hence
\begin{gather}
\int \left[f_{0,1}(\kett)-g_{0,1}(\kett)\right] d\psi  \nonumber \\
\geq-(\delta_{0,\gamma} + \delta_{1,\gamma} ) + \int_{(A_{0,\gamma}\cup A_{1,\gamma})^c} \left[f_{0,1}(\kett)-g_{0,1}(\kett)\right] d\psi,\label{eq:explicit}
\end{gather}
where 
\medskip
\begin{align}
&\mu\left((A_{0,\gamma}\cup A_{1,\gamma})^c\right) \nonumber \\
&\geq \frac{1}{2}+\frac{1}{4\pi}\int_{\frac{\pi}{2}}^{\frac{3\pi}{2}}\cos\left(2\text{arccot}\left(-\frac{\tr(E_0\Pi_1)}{2\gamma\cos(\phi)}\right)\right)\nonumber \\
&\:\: \:\:\:\:\:\: \:+\frac{1}{4\pi}\int_{\frac{\pi}{2}}^{\frac{3\pi}{2}}\cos\left(2\text{arccot}\left(-\frac{\tr(E_1\Pi_0)}{2\gamma\cos(\phi)}\right)\right)
\end{align}
 and, if $\kett \in (A_{0,\gamma}\cup A_{1,\gamma})^c$, 
\begin{equation}
f_{0,1}(\kett)-g_{0,1}(\kett) \geq 0.
\end{equation}
\medskip

This gives the desired sufficient condition for $\int \left[f_{0,1}(\kett)-g_{0,1}(\kett)\right]d\psi \geq 0$, and so a sufficient condition for the lower bound to be valid:
\begin{align}\label{eq:sufficient}
\int_{(A_{0,\gamma}\cup A_{1,\gamma})^c} \left[f_{0,1}(\kett)-g_{0,1}(\kett)\right] d\psi \geq \delta_{0,\gamma} + \delta_{1,\gamma} 
\end{align}
Note that the right hand side of Eq.~(\ref{eq:sufficient}) is typically small and the left hand is non-negative with the integral taken over a set with large measure. 

We have shown here that, for a single qubit, an analytical sufficient condition for the lower bound to be valid does indeed exist. This sufficient condition explicitly defines a regime where our bounds are provably valid. However, clearly it will also be useful to numerically analyze the exact relationship between the average measurement fidelity and our derived lower bound $lb$ for the single-qubit case. This analysis will further aid in shaping our understanding of the validity of the bounds and is performed in the next section. We find that the lower bound is \emph{always} valid, independent of the magnitude of the coherence.

\subsection{Numerical Examples For the Single-Qubit Case}\label{sec:numerics}

In the previous section we obtained a sufficient condition for $lb$ to be valid in the case of a single qubit. We now numerically analyze various single-qubit examples to observe the \emph{exact} behavior of $lb$ relative to the exact value of $\FEM$, which will also help clarify many of the technical details presented thus far. 

In order to perform integration over the Fubini-Study measure on single-qubit states, we again use the Bloch sphere representation of a single qubit state
\begin{equation}
\kett = \cos\left(\frac{\theta}{2}\right) |0\rangle + \sin\left(\frac{\theta}{2}\right) e^{i\phi} |1\rangle,
\end{equation}
where $\theta \in [0,\pi]$, $\phi \in [0,2\pi]$. Thus any function $f$ on pure states can be written as a function $f(\theta,\phi)$ on the unit sphere $\mathbbm{S}^2\subset \mathbbm{R}^3$. In addition, the integral of $f(\theta,\phi)$ with respect to the Fubini-Study measure is just the usual double integral
\begin{equation}
\frac{1}{4\pi}\int_{\theta=0}^{\pi} \int_{\phi=0}^{2\pi}f(\theta,\phi)\sin(\theta)d\theta d\phi,
\end{equation}
where the
normalization $\frac{1}{4\pi}$ ensures we are integrating over a probability measure.

Our goal is to provide a direct comparison between $\FEM$ and $lb$ for the case of a single-qubit projective measurement in the computational basis. As before, we write the noisy POVM elements as $\{E_0,E_1\}$. From Eq.~(\ref{eq:twosums}), we have that the exact value of the average measurement fidelity, $\FEM$, is given by
\begin{gather}
\FEM=\int F(\Eop(\kett\braa),\Meas(\kett\braa))d\psi \nonumber \\
= \sum_{k=0}^1\left[\int r_kp_kd\psi\right] + \sum_{l\neq m}\left[\int\sqrt{r_lr_mp_lp_m}d\psi\right]\nonumber \\
= \sum_{k=0}^1\left[\int r_kp_kd\psi\right] + 2\int\sqrt{r_0r_1p_0p_1}d\psi
\end{gather}
where, using the Bloch sphere representation and the elements of the POVM $\{E_0,E_1\}$, we can write
\begin{align}
p_0&= \cos ^2\left(\frac{\theta}{2}\right), \nonumber \\
p_1&=1-p_0 = \sin^2\left(\frac{\theta}{2}\right), \nonumber \\
r_0&=E_0^{0,0}\cos ^2\left(\frac{\theta}{2}\right) + E_0^{1,1}\sin ^2\left(\frac{\theta}{2}\right) + \text{Re}\left(E_0^{0,1}e^{i\phi}\right)\sin(\theta),\nonumber \\
r_1 &= 1-r_0. \label{eq:singleset}
\end{align}
For each $k = 0,1$ we can compute $\int r_kp_kd\psi$ exactly by the techniques introduced in Sec.~\ref{sec:firstset} (specifically see Eq.~(\ref{eq:exactfirstset})). However as noted in Sec.~\ref{sec:secondset}, calculating $\int\sqrt{r_0r_1p_0p_1}d\psi$ analytically is generally not possible. One can however directly input the expressions in Eq.'s~(\ref{eq:singleset}) into $\int\sqrt{r_0r_1p_0p_1}d\psi$ so that a numerical analysis can be performed.

The POVM elements $\{E_0,E_1\}$ that model the noise must each be positive semidefinite and $E_0+E_1=\Id$. Hence, we write
\begin{gather}
E_0 = \left(\begin{array}{cc}
u_0 & \gamma \\
\gamma^* & 1-u_0
\end{array} \right),
\end{gather}
\begin{gather}
E_1 = \left(\begin{array}{cc}
1-u_0 & -\gamma \\
-\gamma^* & u_0
\end{array} \right),
\end{gather}
where $\gamma \in \mathbbm{C}$ is given by
\begin{equation}
\gamma = re^{i\phi}
\end{equation} 
and $\phi \in [0,2\pi)$. We have that the linear operator $E_0$ ($E_1$) is positive semidefinite if and only if its leading principal minors are non-negative. Since $d=2$, this reduces to the condition
\begin{align}
|\gamma| = r &\leq \sqrt{u_0(1-u_0)}.
\end{align}

For various cases, we observe how the exact measurement fidelity compares to $lb$ as $R$ approaches its maximum value 
\begin{align}
R_{\text{max}}:= \sqrt{u_0(1-u_0)}.
\end{align}
Since the results will only depend on $R$ and not $\phi$, without loss of generality, one can assume $\gamma \in \mathbbm{R}$ so that $\gamma = R$. As a verification of this, the numerics given below were performed for various $\phi \in [0,2\pi)$ and, as expected, the results were independent of $\phi$. 

\begin{figure}\begin{center}
\includegraphics[width=.50\textwidth,height=0.50\textheight]{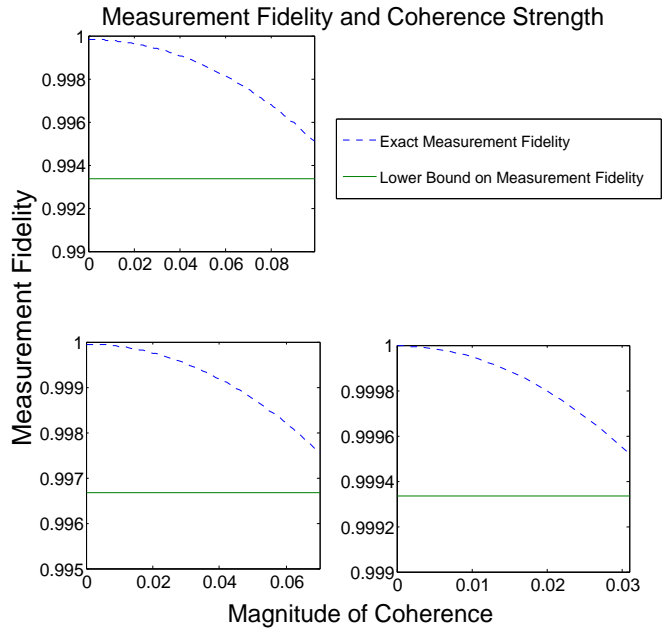}
\caption{\label{Fig:Fid} Plot of exact average measurement fidelity $\FEM$ (blue-dashed) against coherence strength $|\gamma|$ for three different values of $u_0$: $0.99$ (upper left), $0.995$ (bottom left), and $0.999$ (bottom right). The lower bound $lb$ for $\FEM$ in each case is also given (green). The magnitude of the coherence was varied in each case up to its maximum value $R_{\text{max}}$.}
\end{center}
\end{figure}

Three of the cases we analyzed are $u_0=0.99$, $0.995$, and $0.999$. The values of $lb$ and $ub$ for each case are contained in Table~\ref{table:Cases}.
Plots of the average measurement error $\FEM$ as a function of $|\gamma|$ for each case are given in Fig.~\ref{Fig:Fid}. The horizontal axis is cut off at $R_{\text{max}}$ for each case. Thus, all admissible values of the coherence are provided.

\begin{table}[h!]
\caption{Values of $lb$ and $ub$ for 3 Cases of $u_0$.}
\centering
\begin{tabular}{c c c c}
\hline\hline
$u_0$  \:\:\:  & $0.99$ \:\:\: & $0.995$ \:\:\: & $0.999$ \\ [0.9ex] 
\hline
$lb$ \:\:\:  & $0.9933$\:\:\:  & $0.9967$ \:\:\:  & $0.9993$ \\
$ub$ \:\:\:  & $0.0067$ \:\:\:  & $0.0033$ \:\:\:  & $0.0007$ \\ [1ex]
\hline
\end{tabular}
\label{table:Cases}
\end{table}

Various important features are evident from Fig.~\ref{Fig:Fid}. First, $lb$ is \emph{always} a lower bound for $\FEM$ (the curves in the plot do not intersect). This agrees with the discussion in Sec.~(\ref{sec:derivation}) where we noted that $lb$ should be a valid lower bound in large generality (even for large coherence) due to both the anti-correlation of the random variables $\tr(\Pi_0\kett\braa)$ and $\tr(\Pi_1\kett\braa)$, and the measure of the set of states for which $g_{0,1} > f_{0,1}$ being small. Second, as expected, the difference between $lb$ and $\FEM$ grows smaller as the magnitude of the coherence increases. Thus, the bounds become more tight as the coherence in the POVM elements increases.

These results suggest that the bounds are always valid regardless of how large $|\gamma|$ is relative to $u_0$. We tested whether there were any violations when $u_0$ was varied from 0.5 to 0.9999 in increments of $10^{-4}$, and the coherence was varied from its minimum to maximum in each case. We found \emph{no} violations of the lower bound.

\section{Characterizing Measurement Probabilities and Output States for Rank-1 PVM's}\label{sec:outputstates}

Ideally, a quantum measurement produces an output state that can be utilized for further purposes. In this case one would like to characterize the error on both the measurement statistics and output states of the measurement. In this section we briefly outline such a method for rank-1 PVM's and discuss conditions for the lower bound on the measurement fidelity $\FEM$ to be valid. 

From Eq.~(\ref{eq:Measurement}) we have that $\mathcal{M}$ is given by
\begin{eqnarray}
\mathcal{M}(\sigma)&=&\sum_{k=1}^d \Pi_k\sigma\Pi_k = \sum_{k=1}^bp_k\Pi_k,\label{eq:Measurementstates} 
\end{eqnarray}
where
\begin{gather}
p_k=\tr\left(\Pi_k\sigma\right),
\end{gather}
and the output state $\Pi_k$ is \emph{independent} of the input state $\sigma$. Note that this assumption is only valid because the ideal measurement is a rank-1 PVM. We assume the noisy measurement process $\Eop$ is of the form
\begin{eqnarray}
\Eop(\sigma)&=&\sum_{k=1}^br_k\rho_k\label{eq:Measurementrank1} 
\end{eqnarray}
where, as before, 
\begin{equation}
r_k = \tr(E_k\sigma)
\end{equation}
is modeled via a POVM $\{E_k\}_{k=1}^d$. We assume the $\rho_k$ are independent of $\sigma$ since, ideally, the output states $\Pi_k$ are independent of $\sigma$. However, it may be the case that there is some dependence of $\rho_k$ on the input $\sigma$. When the dependence is weak we expect the results presented here to generally still be valid, however the extent to which this is the case is left as an area for further research. When there is large dependence, clearly new techniques will have to be applied. There are simple tests one could perform to help detect such state-dependence of output states. For instance, suppose we take two input states $|\psi_1\rangle$ and $|\psi_2\rangle$. We input $|\psi_1\rangle$ into the measurement and observe whether outcome ``$k$" is obtained. Whenever it is, we can make a successive measurement on the output state $\rho_k(|\psi_1\rangle)$ and observe whether outcome ``$k$" is obtained again. Repeating this many times produces the statistic $\tr(\rho_k(\psi_1)E_k)$. Repeating this procedure for $|\psi_2\rangle$ gives $\tr(\rho_k(\psi_2) E_k)$. If these quantities differ then there must be some input state-dependence in the output states of the measurement. 

The average measurement fidelity is still given by
\begin{eqnarray}
\FEM&=&\int F_{\Eop,\mathcal{M}}(\kett\braa)d\psi \nonumber \\
&=& \int F\left(\sum_kr_k\rho_k,\sum_k p_k\Pi_k\right)d\psi \label{eq:exp1}
\end{eqnarray}
and, as before,
\begin{equation}
\rEM =1- \int F_{\Eop,\mathcal{M}}(\kett\braa) d\psi.\label{eq:rdefstate}
\end{equation}

In this case where we are also interested in output states, the expression for $\FEM$ is much more complicated. However, we can use strong concavity of the fidelity~\cite{Fuc96} to obtain
\begin{gather}
\FEM =\int F(\Eop(\kett),\Meas(\braa)) \nonumber \\
\geq \int \left(\sum_k\sqrt{r_kp_k} \sqrt{\tr\left(\rho_k\Pi_k\right)}\right)^2 d\psi.
\end{gather}
Since $\tr\left(\rho_k\Pi_k\right)$ is just the fidelity between the $k$'th ideal and noisy output states of the measurement, we define
\begin{gather}
F_k:=F(\rho_k,\Pi_k)=\tr\left(\rho_k\Pi_k\right).
\end{gather}
Moreover, since there is only state-dependence in $p_k$ and $r_k$, we have
\begin{gather}
\FEM \geq \sum_{(l,m) \in \mathcal{D}} \left[\int\sqrt{r_lr_mp_lp_m}d\psi\right]\sqrt{F_lF_m}.
\end{gather}
From Sec.~\ref{sec:derivation}, and specifically Eq.~(\ref{eq:secondsetlb}), this gives
\begin{equation}
\FEM \geq  \frac{1+ d\overline{Y}}{1+d},
\end{equation}
where
\begin{equation}\label{eq:Ybardef}
\overline{Y}:=\frac{1}{d^2}\displaystyle{\sum_{(l,m) \in \mathcal{D}}}\sqrt{u_lu_m}\sqrt{F_lF_m}.
\end{equation}
At this point, we could define $lb=\frac{1+ d\overline{Y}}{1+d}$ (and thus $ub=1-\frac{1+ d\overline{Y}}{1+d}$), except the $F_k$ cannot be obtained directly from the noisy measurement $\Eop$. The reason for this is that $F_k$ describes the overlap between the output state from the ideal measurement, $\Pi_k$, and the output state from the real measurement, $\rho_k$.Ê Equivalently, this is the same as the result obtained by making an ideal measurement in the $\Pi_k$ basis on the real measurement output state $\rho_k$.Ê By definition however, we cannot make an ideal measurement $\mathcal{M}$. All we can do is make another real measurement $\Eop$ and look at resulting outcome probabilities, namely $\tr(\rho_k E_k)$.

%
%
%

Let us denote $Q_k$ to be this probability
\begin{align}
Q_k&=\mathbbm{P}\left[\text{obtaining outcome} \: k\:  \text{from} \: \Eop\:  \text{if the input is} \: \rho_k\right]\nonumber \\
&= \tr(\rho_kE_k)
\end{align}
where the notation ``$\mathbbm{P}(\cdot)$" means ``probability of". Thus, \emph{if} it is true that for each $k$
\begin{equation}
F_k=\tr\left(\rho_k\Pi_k\right) \geq Q_k=\tr\left(\rho_kE_k\right),\label{eq:condstate}
\end{equation}
then we have
\begin{align}
\FEM &\geq lb:= \frac{1+ d\overline{Z}}{1+d},  \label{eq:lowerboundstate}
\end{align}
where
\begin{equation}\label{eq:Zbardef}
\overline{Z}:=\frac{1}{d^2}\displaystyle{\sum_{(l,m) \in \mathcal{D}}}\sqrt{u_lu_m}\sqrt{Q_lQ_m}.
\end{equation}

%
%

Motivation for why the assumption in Eq.~(\ref{eq:condstate}) will often be true is that ideally, for each $k$, $E_k=\Pi_k$. Hence, $\tr\left(\rho_k\Pi_k\right)$ is just the $k$'th diagonal element of $\rho_k$, $\rho_k^{k,k}$. Since $E_k \leq \mathbbm{1}$, if $E_k$ is diagonal in the $\{|\psi_k\rangle\}$ basis, then Eq.~(\ref{eq:condstate}) is guaranteed to hold. This is because the probability amplitude of $\Pi_k$ is spread across the diagonal elements of $E_k$. Even when there is coherence in the $E_k$ (non-zero off-diagonal elements) we expect the inequality should still hold in realistic situations, however a more detailed analysis of conditions for this to occur is required. Note also that, realistically, one does not need Eq.~(\ref{eq:condstate}) to hold for every $k$, just a large enough number to ensure $\overline{Y}\geq \overline{Z}$. We now provide the experimental protocol for determining $lb$.



\section{Experimental Protocol: Rank-1 PVM's With Output States}\label{sec:protocolstates}

 \underline{Goal}: Obtain $lb$ in the case of imperfect measurement probabilities \emph{and} output states. 

\bigskip

 \underline{Protocol}:

\medskip


\medskip

\noindent Step 1: Choose a pair of indices $(l,m)$ uniformly at random from $\mathcal{D}=\{0,...,d-1\} \times \{0,...,d-1\}$.

\bigskip

\noindent Step 2: For each $j = l$ and $j=m$,

\bigskip

a) Prepare the quantum state $\Pi_j$, perform the noisy measurement $\Eop$ on $\Pi_j$, and record whether outcome ``$j$" is obtained,

\bigskip

b) If outcome $``j"$ is obtained in a), repeat the measurement on the current state of the system, and record whether outcome ``$j$" is obtained again,

\bigskip 

c) Repeat Steps a) and b) many times and denote the frequency of obtaining $``j"$ in each step by $\hat{u}_j$ and $\hat{Q}_j$ respectively.


\medskip

\noindent (see below for a discussion of the number of repetitions required to estimate $u_j$ and $Q_j$ to desired accuracies).

\bigskip

\noindent Step 3: Repeat Steps 1 and 2 $K$ times, where $K$ is dictated by the desired accuracy and confidence in estimating $lb$ 

\medskip

\noindent (see below for a discussion of the size of $K$).


\bigskip

\noindent Step 4: Compute the estimator, $\hat{lb}$, of $lb$ (where $lb$ is given in Eq.~(\ref{eq:lowerboundstate})) to accuracy and confidence dictated by the $K$ trials $\{(1_1,1_2),...,(K_1,K_2)\}$ from Step 3

\begin{align}
\hat{lb}
&= \frac{1+d\left[\frac{1}{K}\displaystyle{\sum_{(k_1,k_2)}\sqrt{u_{k_1}u_{k_2}}}\sqrt{Q_{k_1}Q_{k_2}}\right]}{1+d}.
\end{align}

\bigskip

This concludes the protocol.

\bigskip

We should emphasize various points about this protocol. First, similar to the case of no output states (see Sec~\ref{sec:protocol}), the number of trials required to implement the above protocol is independent of $d$, and only depends on the desired accuracy and confidence of the estimates in each of Steps 2 c) and 3. See Sec.'s~\ref{sec:trials1} and~\ref{sec:trials2} for respective discussions about the time-complexity of each of these steps. 

Second, $lb$ can be estimated using only:
\begin{enumerate}
\item Sequential applications of the noisy measurement and,
\item The ability to prepare each of the $d$ pure input states $\Pi_j$. 
\end{enumerate}
Lastly, it is straightforward to show that the two necessary conditions given previously in Sec.~\ref{sec:necessary} for $lb$ to be a useful lower bound also hold here. Indeed,
\begin{enumerate}
\item In the limit of $\FEM \uparrow 1$,
\begin{equation}
lb \uparrow 1,
\end{equation}
and
\item $lb$ scales well in $d$. 
\end{enumerate}
The first condition holds since, as $\FEM \uparrow 1$, it must be the case that $E_k \rightarrow \Pi_k$ and $\rho_k \rightarrow \Pi_k$. Therefore, for every $k$, $u_k \rightarrow 1$ and $Q_k \rightarrow 1$. The second condition holds using an analogous argument as that given in Sec.~\ref{sec:scaling}

\section{Resource Analysis}\label{sec:resources}

In this section we discuss the time-complexity and resources required for the protocols and compare the time-complexity of the protocol with a full reconstruction of the noisy measurement (ie. a full reconstruction of the POVM elements $E_k$). First, we analyze the number of trials required in Step 2b) of the protocol in Sec.~\ref{sec:protocol} and Step 2c) of the protocol in Sec.~\ref{sec:protocolstates}. Afterwards, we analyze the number of trials required in Step 3 of both protocols.

\subsection{Number of Trials Required in Step 2b) of Sec.~\ref{sec:protocol} and Step 2c) of Sec.~\ref{sec:protocolstates}} \label{sec:trials1}

We first explicitly analyze the number of trials for Step 2b) of Sec.~\ref{sec:protocol}. The discussion carries over in a straightforward manner to Step 2c) of Sec.~\ref{sec:protocolstates}.

\subsubsection{Step 2b) of Sec.\ref{sec:protocol}}

For each $k \in \{0,1,...,d-1\}$, we would like to understand how many samples are required to obtain an estimate of $u_k$. Let us fix $k$ and define $z_k=\tr(\Pi_k(\Id-E_k))$ where, since $\{E_k\}_{k=1}^d$ is a POVM,
\begin{equation}
\Id-E_k=\sum_{j\neq k}^dE_j.
\end{equation}
Then
\begin{eqnarray}
z_k &\geq& 0, \nonumber \\
u_k+z_k &=& \tr(\Pi_k) =1,
\end{eqnarray}
so $\{u_k,z_k\}=\{u_k,1-u_k\}$ forms a probability distribution on the binary measurement outcome space $\{``k", ``\text{not} \: k"\}$. Let us define the random variable $V$ which takes the values $\{``k", ``\text{not} \: k"\}$ with associated probabilities $\{u_k,1-u_k\}$. Then, $V$ is an asymmetric Bernoulli random variable~\cite{Schervish} (it is asymmetric since in general $u_k \neq \frac{1}{2}$), and we can call outcome $``k"$ a success and outcome $``\text{not} \: k"$ a failure.

Let $\hat{u}_k$ be the estimator of $u_k$ obtained in 2b of the protocol. Since $u_k$ is just the probability of success in a Bernoulli random variable, we can use maximum likelihood estimation (MLE) techniques~\cite{Schervish} to determine the number of trials required to estimate $u_k$ to some desired accuracy and confidence. As the trials performed in the protocol are independent, standard MLE gives that $\hat{u}_k$ is just equal to the frequency of obtaining outcome ``$k$". Thus, if $N$ is the number of trials and $n_k$ is the number of times outcome ``$k$" is observed, standard MLE gives the estimate
\begin{gather}
\hat{u}_k=\frac{n_k}{N}
\end{gather}
which agrees with intuition.

There can be a problem with this procedure if $u_k$ is very close to 1. In particular, if ``$k$" is observed in all $N$ trials then a probability of 1 is assigned to ``$k$" and a probability of 0 is assigned to ``not $k$", which is not a physically realistic scenario. This is an example of a much more general problem that can arise when using MLE to assign probabilities to rare events that are not observed because the data set is not large enough. One solution to this problem is to utilize additive (Laplace) smoothing methods to augment the MLE procedure~\cite{MRS}. The main idea behind such a smoothing technique is to assign higher (non-zero) weight to low probability outcomes. In our case, the low probability outcome is ``not $k$" and the weight can be controlled by a parameter $\lambda$ according to the formula
\begin{equation}
\hat{u}_k=\frac{n_k + \lambda}{N + 2\lambda}.
\end{equation}
The factor of 2 multiplying $\lambda$ in the denominator arises because there are two possible outcomes of the experiment, and is required to ensure $\{\hat{u}_k,1-\hat{u}_k\}$ is a probability distribution. Note that when this smoothing technique is used, it will tend to fairly bias the estimate of $u_k$ to be smaller than the actual value of $u_k$, so the estimation will be honest.

The key point in terms of time complexity, which we now prove, is that the number of trials $N$ required to estimate $u_k$ to accuracy $\epsilon$ with confidence $1-\delta$ is \emph{independent} of the size of the system. Since the set of possible probability distributions $\{u_k,z_k\}$ satisfies certain consistency conditions, the estimator $\hat{u}_k$ converges in distribution to $u_k$,
\begin{equation}
\hat{u}_k \xrightarrow{D} u_k.
\end{equation}
Moreover, the mean of $\hat{u}_k$ is equal to $u_k$ and the variance of $\hat{u}_k$ scales as 
\begin{equation}
\text{Var}\left(\hat{u}_k\right) \sim \frac{1}{N I(\{u_k,z_k\})},\label{eq:Variance}
\end{equation}
where $I(\{u_k,z_k\})$ is the Fisher information~\cite{Schervish} of one observation of the true probability distribution. The Fisher information contained in one observation is a measure of how much information about $u_k$ is gained on average from observing $\{``k", ``\text{not} \: k"\}$ (with distribution $\{u_k,z_k\})$. If $u_k \sim 1$ then the amount of information gained on average is large. 

Since $\{``k", ``\text{not} \: k"\}$ is a Bernoulli random variable, it is possible to explicitly compute the Fisher information of one observation~\cite{Schervish},
\begin{equation}
I(\{u_k,z_k\}) = \frac{1}{u_kz_k} = \frac{1}{u_k(1-u_k)}.\label{eq:Fisherone}
\end{equation}
Hence, from Eq.'s~(\ref{eq:Variance}) and (\ref{eq:Fisherone}), we have that 
\begin{equation}
\text{Var}\left(\hat{u}_k\right) \sim \frac{u_k(1-u_k)}{N}.
\end{equation}
Now, suppose we want to estimate $u_k$ to accuracy $\epsilon$ with confidence $1-\delta$, ie. we want
\begin{equation}
\mathbb{P}\left(|\hat{u}_k-u_k| \geq \epsilon\right)\leq \delta.
\end{equation}
By Chebyshev's theorem we have for any integer $j \geq 1$,
\begin{equation}
\mathbb{P}\left(|\hat{u}_k-u_k| \geq j\sigma(\hat{u}_k)\right)\leq\frac{1}{j^2}.
\end{equation}
Choose $j_\delta$ to be the smallest $j$ such that 
\begin{equation}
\frac{1}{j_\delta^2}\leq \delta.
\end{equation}
Then we have
\begin{equation}
\mathbb{P}\left(|\hat{u}_k-u_k| \geq \frac{j_\delta\sqrt{u_k(1-u_k)}}{\sqrt{N}}\right)\leq\frac{1}{j_\delta^2},
\end{equation}
and so we set
\begin{equation}
\frac{j_\delta\sqrt{u_k(1-u_k)}}{\sqrt{N}} \leq \epsilon.
\end{equation}
This gives
\begin{equation}
N \geq \frac{j_\delta^2u_k(1-u_k)}{\epsilon^2}.
\end{equation}
In total, if $N \geq \frac{j_\delta^2u_k(1-u_k)}{\epsilon^2}$ with $j_\delta \geq \frac{1}{\delta}$, then
\begin{equation}
\mathbb{P}\left(|\hat{u}_k-u_k| \geq \epsilon\right)\leq \delta.
\end{equation}
Under the assumption that the $u_k$ are independent of $d$, we have that $N$ is independent of $d$. Hence, the number of trials required to estimate $u_k$ to accuracy $\epsilon$ and confidence $1-\delta$ is independent of $d$. If the probabilities $u_k$ do depend on $d$ then the number of trials will be a function of $d$. Determining the extent to which the $u_k$ can have dependence on $d$ is an interesting question and likely depends on the particular scenario at hand. This is left as a topic for further investigation.


\subsubsection{Step 2c) of Sec.~\ref{sec:protocolstates}}

 In the protocol of Sec.~\ref{sec:protocolstates} we have to estimate both $u_k$ and $Q_k$ where,
$u_k$ is the probability of obtaining outcome ``$k$" on the first measurement and $Q_k$ is the probability of obtaining outcome ``$k$" on a second measurement \emph{given} the result of the first measurement is $``k"$. Note that observing ``$k$" in the second measurement is also a Bernoulli random variable, so the discussion from above carries over in an analogous manner to estimating $Q_k$. 

Let $N_1$ and $N_2$ be the number of trials required to estimate $u_k$ and $Q_k$ to each of their respective accuracies and confidences. Suppose we perform the first measurement $N_1$ times (so we have estimated $u_k$ to its desired accuracy and confidence). If the number of times ``$k$" is observed over these trials is greater than $N_2$ then we have estimated both $u_k$ and $Q_k$ to their desired accuracies. If the number of times ``$k$" is observed in the first measurement is less than $N_2$ then we keep repeating until $N_2$ outcomes of ``$k$" are recorded in the first measurement. Let $M_2$ be the number of times the first measurement has to be performed before $N_2$ values of ``$k$" are recorded. The total number of trials is no more than 
\begin{align}
N&=\text{min}\{N_1,M_2\}.
\end{align}
As $N$ only depends on the accuracies, confidences, and values of $u_k$ and $Q_k$, it is independent of $d$.

%

\subsection{Number of Trials Required in Step 3 of Sec.~(\ref{sec:protocol}) and Step 3 of Sec.~(\ref{sec:protocolstates})}\label{sec:trials2}

Let us now discuss how many trials are required in Step 3 of each of the protocols. The argument is the same for each protocol so, without loss of generality, we use the notation from Sec.~\ref{sec:protocol} (the discussion for Sec.~(\ref{sec:protocolstates}) follows by replacing $X$ with $Z$ and making appropriate changes). The random variable $X:\mathcal{D}\rightarrow [0,1]$ is defined by
\begin{equation}
X(l,m)=\sqrt{u_lu_m}
\end{equation}
(for Sec.~\ref{sec:protocolstates} this will be $Z(l,m)=\sqrt{u_lu_m}\sqrt{Q_lQ_m}$) and the expectation value of $X$ is denoted $\overline{X}$. Suppose one wants to estimate $\overline{X}$ to accuracy $\epsilon$ and confidence $1-\delta$. Let $\hat{\overline{X}}$ denote the estimator of $\overline{X}$ obtained from taking the average of $K$ independent samples, $\{X_1,...,X_K\}$, of $X$
\begin{align}
\hat{\overline{X}} &=  \frac{1}{K}\displaystyle{\sum_{(k_1,k_2)}\sqrt{u_{k_1}u_{k_2}}}.
\end{align}
By Hoeffding's inequality
\begin{eqnarray}
\mathbbm{P}\left(\left|\hat{\overline{X}}-\overline{X}\right| \geq \epsilon \right) &\leq& 2 e^{\frac{-2(k\epsilon)^2}{K\left(b-a\right)^2}} \nonumber \\
&=& 2 e^{\frac{-2K\epsilon^2}{\left(b-a\right)^2}}
\end{eqnarray}
where $[a,b]$ is the range of $X$ (here $[a,b]\subseteq [0,1]$). Hence, setting 
\begin{equation}
\delta=2 e^{\frac{-2K\epsilon^2}{\left(b-a\right)^2}},
\end{equation}
gives
\begin{equation} \label{eq:notrials}
K=\frac{\ln \left(\frac{2}{\delta}\right)(b-a)^2}{2\epsilon ^2} \leq \frac{\ln \left(\frac{2}{\delta}\right)}{2\epsilon ^2} 
\end{equation}
\noindent which is independent of $d$. In practice, $b-a \ll 1$ which will improve this bound on $K$.

\subsection{Complete Reconstruction of POVM Elements}\label{sec:reconstruction}

For completeness, we provide an explicit protocol and determination of the time-complexity required to perform complete tomography of the noisy POVM elements $\{E_k\}_{k=1}^d$. The idea is to input various pure states into the noisy measurement and analyzing the frequency of obtaining particular outcomes.  If the matrix representation of $E_k$ in the $|\psi_m\rangle$ basis is written as $E_k^{i,j}$, then there are $\frac{d^2+d}{2}$ elements on the main diagonal and upper triangular section of $E_k$ that need to be estimated (since each $E_k \geq 0$, the lower triangular part of $E_k$ is completely determined by the upper triangular part). One can estimate the elements of $E_k$ by first defining the pure states
\begin{align}
|\psi_{i,j}^+\rangle &= \frac{|\psi_i\rangle+|\psi_j\rangle}{\sqrt{2}} \:\:\: \text{and}\:\:\: |\psi_{i,j}^-\rangle = \frac{|\psi_i\rangle+i|\psi_j\rangle}{\sqrt{2}}.\nonumber
\end{align}
Then, since
\begin{align}
|\psi_i\rangle\langle \psi_j|&= |\psi_{i,j}^+\rangle\langle \psi_{i,j}^+| + i|\psi_{i,j}^-\rangle\langle \psi_{i,j}^-|\nonumber \\
&\:\:\: -\left(\frac{1+i}{2}\right)|\psi_i\rangle\langle\psi_i|-\left(\frac{1+i}{2}\right)|\psi_j\rangle\langle\psi_j|,
\end{align}
we have
\begin{align}
E_k^{i,j}&=\tr\left(E_k\Pi_{i,j}^+\right) + i\tr\left(E_k\Pi_{i,j}^-\right) \nonumber \\
&\:\:\: - \left(\frac{1+i}{2}\right)\tr(E_k\Pi_i) - \left(\frac{1+i}{2}\right)\tr(E_k\Pi_j).\label{eq:offdiags}
\end{align}
\noindent The algorithm to determine the set $\{E_k\}_{k=1}^d$ is as follows. 

\bigskip

Step 1: For each state $|\psi_{j,j}\rangle \in \{|\psi_{k,k}\rangle\}_{k=1}^d$, input $|\psi_{j,j}\rangle$ into the noisy measurement $\Eop$ many times and record the frequency of obtaining \emph{each} of the $d$ different possible outcomes ``$k$". For each $k$ this gives $\tr(E_k|\psi_{j,j}\rangle\langle \psi_{j,j}|)$. 

\bigskip

Step 2: For each state $|\psi_{i,j}^+\rangle \in \{|\psi_{k,l}^+\rangle\}_{k,l=1}^d$, input $|\psi_{i,j}^+\rangle$ into the noisy measurement $\Eop$ many times and record the frequency of obtaining \emph{each} of the $d$ different possible outcomes ``$k$". For each $k$ this gives $\tr(E_k|\psi_{i,j}^+\rangle\langle \psi_{i,j}^+|)=\tr\left(E_k\Pi_{i,j}^+\right)$. 

\bigskip

Step 3: For each state $|\psi_{i,j}^-\rangle \in \{|\psi_{k,l}^-\rangle\}_{k,l=1}^d$, input $|\psi_{i,j}^-\rangle$ into the noisy measurement $\Eop$ many times and record the frequency of obtaining \emph{each} of the $d$ different possible outcomes ``$k$". For each $k$ this gives $\tr(E_k|\psi_{i,j}^-\rangle\langle \psi_{i,j}^-|)=\tr\left(E_k\Pi_{i,j}^-\right)$. 

\bigskip

Step 4: Combine all of the elements estimated in Steps 1 through 3 to construct the $E_k$. Step 1 gives the diagonal elements of the $E_k$ since
\begin{equation}
E_k^{j,j}=\tr(E_k|\psi_{j,j}\rangle\langle \psi_{j,j}|).
\end{equation}
Eq.~(\ref{eq:offdiags}) and Steps 1 through 3 give all of the off-diagonal elements $E_k^{i,j}$. This concludes the protocol.

\bigskip

The number of trials one will have to perform is again dictated by MLE. It is important to note that the MLE procedure in this case is more involved than the simple Bernoulli procedure for our protocol (described in Sec.~\ref{sec:trials1}). This is because, when the noisy measurement is performed, one must keep track of which value of $j$ is obtained (not just whether the outcome was $j$ or not).  Thus the number of trials in each step will be greater than that required to estimate the $u_k$ because events with small probability may rarely be seen (if at all). Smoothing techniques will likely have to be employed to ensure rare events are not assigned zero probability.
Thus, the determination of each $\tr(E_k|\phi\rangle\langle \phi|)$, where $|\phi\rangle$ is one of $|\psi_{j,j}\rangle$, $|\psi_{i,j}^+\rangle$, or $|\psi_{i,j}^-\rangle$, requires greater time-complexity than that of estimating each $u_k$. Since there are $d + 2\frac{d(d-1)}{2} = d^2$ such $|\phi\rangle$, one will have to estimate $d^3$ different probabilities over Steps 1 through 3 using MLE. 
Hence, the full reconstruction requires the estimation of $d^3$ probabilities with more complicated post-processing of the measurement data (as well as a larger number of trials to estimate each probability). In addition, there is an added complexity in preparing $d^3$ different input states since more complex rotations may be required.

\section{Discussion}\label{sec:discussion}

We have provided a straightforward, efficient, and experimentally implementable method for obtaining estimates for lower bounds on the average fidelity of projective (rank-1) quantum measurements. As realizations of quantum protocols scale to larger sizes, and full measurement tomography becomes impossible to implement, our protocol can potentially be used as a simple method to benchmark the performance of a measuring device. 

We have discussed conditions for the validity of the bounds and explained why they should hold in extremely general situations. 
The bounds could also potentially be useful as estimates of the average measurement fidelity in the small-error regime. We have presented a set of numerical examples for a single-qubit system. In every instance analyzed, that is, for all possible values of the coherence in the noisy POVM operators, the bounds were found to be valid. This provides further evidence that the bounds should hold in extremely general situations and should be useful in practice. In addition, the bounds became better approximations of the actual measurement fidelity as the magnitude of the coherence increased. 
 
The protocols are scalable and only require the ability to prepare states from a basis set and perform the noisy measurement (sequentially when there are output states) to estimate $d$ probabilities. In addition, post-processing of the data is completely straightforward and avoids the difficulties in associating large sets of tomographic data to valid mathematical objects. This can be compared with a full reconstruction of the POVM elements of the noisy measurement, which requires the preparation of $d^2$ input states, the estimation of $d^3$ probabilities, and more involved post-processing. In addition, the $d^2$ input states required for tomography can be highly complex. In many situations, these input states will have to be prepared using complicated unitary rotations. Minimizing the required set of input states, as well as their complexity, is important for obtaining a more faithful characterization of the measurement that is less prone to state-preparation errors.

It is important to note that the experimental protocols provide an \emph{estimate} for a lower bound on the average measurement fidelity. More precisely, one chooses a number of trials $K$ to obtain the estimate, where $K$ depends on the desired accuracy $\epsilon$ and confidence $1-\delta$ of the estimate. Thus, under the assumption that the lower bound is valid, the true value of the average measurement fidelity is no more than $\epsilon$ units of distance \emph{below} the estimated lower bound with confidence $1-\delta$. 

There are a number of different questions and avenues for future research. First, we have focused on the case of rank-1 PVM's, however we expect that 
our results can be extended to higher-rank PVM's, especially low-rank PVM's in large Hilbert spaces. As well, since any POVM can be implemented via a PVM on an extended Hilbert space, the protocol can potentially give information regarding the quality of implementations of POVM measurements. It will also be useful to analyze the extent to which the ideas presented here can be used to characterize non-ideal POVM measurements that are not implemented as PVM's on a larger space. 

While we have shown the bounds derived here should hold in large generality, a deeper understanding of the validity of the bounds will clearly be useful. In most physically relevant cases, where the coherence is not overwhelmingly large, the protocol should produce valid bounds. In any noise estimation or characterization scheme, there is a trade-off between the amount of information one is able to extract and the amount of resources required to implement the scheme. Here, we obtain a single parameter which serves as an upper (lower) bound on the error (fidelity) of the measurement in a scalable amount of time. Ideally, one would like as much information about the physical measurement as possible. Further analysis of schemes that give more information about an imperfect measurement than average fidelities or errors, while retaining properties such as scalability, will clearly be useful.

As previously mentioned, the algorithms given here require the ability to prepare the basis states $\{|\psi_k\rangle\}$ which constitute the measurement. In practice, these states typically have errors and may actually be created by a measurement procedure, which is exactly what we want to characterize. There are various systems however where state preparation is very different from the measurement procedure. For instance, in superconducting qubit systems~\cite{BVJED,KYG}, the system is initialized to the ground state by cooling the system to extremely low temperatures. Coupling the system to a superconducting resonator in a circuit-QED set-up~\cite{BHW,KYG} allows one to perform both state preparation (by applying unitary rotations) and measurements. Typically, unitary rotations have much higher fidelities than measurements and so one expects state preparation to be much more accurate than measurements, which is ideal for the protocols presented in this paper.

We note that running the protocol with noisy states can still provide valid bounds. For instance, if the noisy input states are given by a convex combination of elements of $\{|\psi_k\rangle\}$ then the bounds still hold. The same is also true for noisy input states with small coherence in the $\{|\psi_k\rangle\}$ basis. The performance of the bounds under more general noise models on input states is a topic for future research.



The scalable protocols presented here can be useful for determining the quality of experimental quantum measurements. There is still much to investigate with regard to useful metrics for comparing measurements and proposing experimentally efficient methods for characterizing measurement devices. As experimental quantum systems scale to larger sizes, such methods will be useful for characterizing and controlling multi-qubit systems. 


\begin{acknowledgements}
E.M. acknowledges financial support from the National Science Foundation through grant NSF PHY-1125846. The authors are grateful for helpful discussions with Alexandre Cooper, Joseph Emerson, Jay Gambetta, Masoud Mohseni, and Marcus Silva.
\end{acknowledgements}

%
%
%

\begin{thebibliography}{10}

\bibitem{Fey82}
R.~Feynman, Internat. J. Theoret. Phys \textbf{21}, 6 (1982).

\bibitem{GLM}
V.~Giovannetti, S.~Lloyd, and L.~Maccone, Science \textbf{306}, 1330 (2004).

\bibitem{Sho94}
P.~Shor, in \emph{Proceedings of the 35'th Annual Symposium on Foundations of
  Computer Science (FOCS)} (IEEE Press, Los Alamitos, CA, 1994), pp. 124--134.

\bibitem{Llo96}
S.~Lloyd, Science \textbf{273}, 1073 (1996).

\bibitem{Deutsch85}
D.~Deutsch, Proc. Roy. Soc. Lond. A \textbf{400}, 97 (1985).

\bibitem{RB01}
R.~Raussendorf and H.~Briegel, Phys. Rev. Lett. \textbf{86}, 5188 (2001).

\bibitem{Kit97-2}
A.~Kitaev, Annals of Physics \textbf{303}, 2 (1997).

\bibitem{BVJED}
V.~Bouchiat et~al., Phys. Scr. A \textbf{T76}, 165 (1998).

\bibitem{KYG}
J.~Koch et~al., Phys. Rev. A \textbf{76}, 042319 (2007).

\bibitem{WJ06}
J.~Wrachtrup and F.~Jelezko, Journal of Physics: Condensed Matter \textbf{18},
  S807 (2006).

\bibitem{CZ}
I.~Cirac and P.~Zoller, Phys. Rev. Lett. \textbf{74}, 4091 (1995).

\bibitem{CFH}
D.~Cory, A.~Fahmy, and T.~Havel, in \emph{Proceedings of the 4th Workshop on
  Physics and Computation} (Boston, MA, 1996).

\bibitem{LD}
D.~Loss and D.~Divincenzo, Phys. Rev. A \textbf{57}, 120 (1998).

\bibitem{KLM}
E.~Knill, R.~Laflamme, and G.~J. Milburn, Nature \textbf{409}, 46 (2001).

\bibitem{CN}
I.~Chuang and M.~Nielsen, J. Mod. Opt. \textbf{44}, 2455 (1997).

\bibitem{ABJ}
J.~B. Altepeter et~al., Phys. Rev. Lett. \textbf{90}, 193601 (2003).

\bibitem{ML06}
M.~Mohseni and D.~A. Lidar, Phys. Rev. Lett. \textbf{97}, 170501 (2006).

\bibitem{LFC}
J.~S. Lundeen et~al., Nature Physics \textbf{5}, 27 (2009).

\bibitem{ZCD}
L.~Zhang et~al., Nature Photonics \textbf{6}, 364 (2012).

\bibitem{MGS}
S.~Merkel et~al., \emph{Self-consistent quantum process tomography} (2012),
  arXiv:quant-ph/1211.0322.

\bibitem{MSB}
T.~Monz et~al., Phys. Rev. Lett. \textbf{106}, 130506 (2011).

\bibitem{Sho96}
P.~Shor, in \emph{Proceedings of the 37'th Annual Symposium on Foundations of
  Computer Science (FOCS)} (IEEE Press, Burlington, VT, 1996).

\bibitem{AB-O}
D.~Aharonov and M.~Ben-Or, in \emph{Proceedings of the 29th Annual ACM
  Symposium on Theory of Computing (STOC)} (1997).

\bibitem{KLZ}
E.~Knill, R.~Laflamme, and W.~Zurek, Proc. R. Soc. Lond. A \textbf{454}, 365
  (1997).

\bibitem{Pre97}
J.~Preskill, \emph{Fault tolerant quantum computation} (1997),
  arXiv:quant-ph/9712048.

\bibitem{ESMR}
J.~Emerson et~al., Science \textbf{317}, 1893 (2007).

\bibitem{SMKE}
M.~Silva et~al., Phys. Rev. A \textbf{78}, 012347 (2008).

\bibitem{KLRB}
E.~Knill et~al., Phys. Rev. A \textbf{77}, 012307 (2008).

\bibitem{MGE}
E.~Magesan, J.~M. Gambetta, and J.~Emerson, Phys. Rev. Lett. \textbf{106},
  180504 (2011).

\bibitem{MGJ}
E.~Magesan et~al., Phys. Rev. Lett. \textbf{109}, 080505 (2012).

\bibitem{MSRL}
O.~Moussa et~al., Phys. Rev. Lett. \textbf{109}, 070504 (2012).

\bibitem{BPP}
A.~Bendersky, F.~Pastawski, and J.~Paz, Phys. Rev. Lett. \textbf{100}, 190403
  (2008).

\bibitem{SLCP}
M.~P. da~Silva, O.~Landon-Cardinal, and D.~Poulin, Phys. Rev. Lett.
  \textbf{107}, 210404 (2011).

\bibitem{SBLP}
C.~Schmiegelow et~al., Phys. Rev. Lett. \textbf{107}, 100502 (2011).

\bibitem{FL}
S.~Flammia and Y.-K. Liu, Phys. Rev. Lett. \textbf{106}, 230501 (2011).

\bibitem{Nie02}
M.~Nielsen, Phys. Lett. A \textbf{303}, 249 (2002).

\bibitem{EAZ}
J.~Emerson, R.~Alicki, and K.~Zyczkowski, J. Opt. B: Quantum and Semiclassical
  Optics \textbf{7}, S347 (2005).

\bibitem{MBKE}
E.~Magesan, R.~Blume-Kohout, and J.~Emerson, Phys. Rev. A \textbf{84}, 012309
  (2011).

\bibitem{OC}
O.~Oreshkov and J.~Calsamiglia, Phys. Rev. A \textbf{79}, 032336 (2009).

\bibitem{Paul}
V.~Paulsen, \emph{Completely Bounded Maps and Operator Algebras}, vol.~78
  (Cambridge University Press, UK, 2002).

\bibitem{Got97}
D.~Gottesman, \emph{Stabilizer codes and quantum error correction} (1997),
  ph.D. Thesis, arXiv:quant-ph/9705052.

\bibitem{Got99}
Chaos, Solitons and Fractals \textbf{10}, 1749  (1999).

\bibitem{RDN}
M.~D. Reed et~al., Nature \textbf{482}, 382 (2012).

\bibitem{DS12}
D.~DiVincenzo and F.~Solgun (2012), arXiv:1205.1910.

\bibitem{GLN}
A.~Gilchrist, N.~Langford, and M.~Nielsen, Phys. Rev. A \textbf{71}, 062310
  (2005).

\bibitem{BZ}
I.~Bengtsson and K.~Zyczkowski, \emph{Geometry of Quantum States: An
  Introduction to Quantum Entanglement} (Cambridge University Press, Cambridge,
  UK, 2006).

\bibitem{FvdG}
C.~Fuchs and J.~van~de Graaf, IEEE Trans. Inf. Th. \textbf{45}, 1216 (1999).

\bibitem{Kit97}
A.~Kitaev, Russian Mathematical Surveys \textbf{52}, 1191 (1997).

\bibitem{BK11}
S.~Beigi and R.~Koenig, New J. Phys. \textbf{13}, 093036 (2011).

\bibitem{RBSC}
J.~Renes et~al., J. Math. Phys. \textbf{45}, 2171 (2004).

\bibitem{HJ}
R.~Horn and C.~Johnson, \emph{Matrix Analysis} (Cambridge University Press,
  Cambridge, UK, 1990).

\bibitem{Led01}
M.~Ledoux, \emph{The Concentration of Measure Phenomenon} (American
  Mathematical Society, 2001).

\bibitem{Fuc96}
C.~A. Fuchs, Ph.D. thesis, University of New Mexico (2006), preprint quant-ph
  9601020.

\bibitem{Schervish}
M.~J. Schervish, \emph{Theory of Statistics} (Springer, New York, USA, 1997),
  1st ed.

\bibitem{MRS}
C.~Manning, P.~Raghavan, and H.~Sch{\"u}tze, \emph{Introduction to Information
  Retrieval} (Cambridge University Press, New York, NY, 2008).

\bibitem{BHW}
A.~Blais et~al., Phys. Rev. A \textbf{69}, 062320 (2004).

\end{thebibliography}
%
%
%
%
%

\end{document}